\documentclass[structabstract]{aa} 
\usepackage{booktabs}
\usepackage{amsmath}
\usepackage{graphicx}
\usepackage{txfonts}
\usepackage{natbib}
\usepackage{longtable}
\usepackage{lscape}
\usepackage{multirow}
\usepackage{pdfpages}

\usepackage[colorlinks=true, allcolors=blue]{hyperref}

\bibpunct{(}{)}{;}{a}{}{,}

\newcommand{\kms}{km\,s$^{-1}$}

\newcommand{\degree}{$^{\circ}$}

\begin{document}
\title{Protonated hydrogen cyanide as a tracer of pristine molecular gas}

\author{Y.~Gong\inst{1}, F.~J.~Du\inst{2,3}, C.~Henkel\inst{1,4,5}, A.~M.~Jacob\inst{6,1}, A.~Belloche\inst{1}, J.~Z.~Wang\inst{7}, K.~M.~Menten\inst{1}, W.~Yang\inst{1}, D.~H.~Quan\inst{8,5}, C.~T.~Bop\inst{9},  G.~N.~Ortiz-Le{\'o}n\inst{10}, X.~D.~Tang\inst{5, 11, 12}, M.~R.~Rugel\inst{13,14,1}, S.~Liu\inst{15}}
\offprints{Y. Gong, \email{ygong@mpifr-bonn.mpg.de}}

\institute{
Max-Planck-Institut f{\"u}r Radioastronomie, Auf dem H{\"u}gel 69, D-53121 Bonn, Germany
\and 
Purple Mountain Observatory, Chinese Academy of Sciences, Nanjing 210023, China
\and 
School of Astronomy and Space Science, University of Science and Technology of China, Hefei 230026, China
\and
Astronomy Department, Faculty of Science, King Abdulaziz University, PO Box 80203, Jeddah 21589, Saudi Arabia
\and 
Xinjiang Astronomical Observatory, Chinese Academy of Sciences, 150 Science 1-Street, Urumqi, Xinjiang 830011, China
\and 
William H. Miller III Department of Physics \& Astronomy, Johns Hopkins University, 3400 North Charles Street, Baltimore, MD 21218, USA
\and
Guangxi Key Laboratory for Relativistic Astrophysics, Department of Physics, Guangxi University, Nanning 530004, PR China
\and 
Research Center for Intelligent Computing Platforms, Zhejiang Laboratory, Hangzhou 311100, China
\and 
Univ Rennes, CNRS, IPR (Institut de Physique de Rennes) - UMR 6251, F-35000 Rennes, France
\and 
Instituto Nacional de Astrof\'isica, \'Optica y Electr\'onica, Apartado Postal 51 y 216, 72000, Puebla, Mexico
\and 
Key Laboratory of Radio Astronomy, Chinese Academy of Sciences, Urumqi 830011, PR China
\and
University of Chinese Academy of Sciences, Beijing 100049, PR China
\and 
Center for Astrophysics $\mid$ Harvard \& Smithsonian, 60  Garden St., Cambridge, MA 02138, USA
\and
National Radio Astronomy Observatory, 1003 Lopezville RD, Socorro, NM 87801, USA
\and 
National Astronomical Observatories, Chinese Academy of Sciences, Beijing 100101, PR China
}

\date{Received date ; accepted date}

\abstract
{Protonated hydrogen cyanide, HCNH$^{+}$, plays a fundamental role in astrochemistry because it is an intermediary in gas-phase ion-neutral reactions within cold molecular clouds. However, the impact of the environment on the chemistry of HCNH$^{+}$ remains poorly understood.}
{We aim to study HCNH$^{+}$, HCN, and HNC, as well as two other chemically related ions, HCO$^{+}$ and N$_{2}$H$^{+}$, in different star formation regions in order to investigate how the environment influences the chemistry of HCNH$^{+}$.}
{With the IRAM-30 m and APEX-12 m telescopes, we carried out HCNH$^{+}$, H$^{13}$CN, HN$^{13}$C, H$^{13}$CO$^{+}$, and N$_{2}$H$^{+}$ imaging observations toward two dark clouds, the  Serpens filament and Serpens South, both of which harbor sites of star formation including protostellar objects as well as regions that are quiescent.}
{We report the first robust distribution of HCNH$^{+}$ in the Serpens filament  and in Serpens South. Our data suggest that HCNH$^{+}$ is abundant in cold and quiescent regions, but is deficient in active star-forming regions. The observed HCNH$^{+}$ fractional abundances relative to H$_{2}$ range from $3.1\times 10^{-11}$ in protostellar cores to $5.9\times 10^{-10}$ in prestellar cores, and the HCNH$^{+}$ abundance generally decreases with increasing H$_{2}$ column density, which suggests that HCNH$^{+}$ coevolves with cloud cores. Our observations and modeling results suggest that the abundance of HCNH$^{+}$ in cold molecular clouds is strongly dependent on the H$_{2}$ number density. The decrease in the abundance of HCNH$^{+}$ is caused by the fact that its main precursors (e.g., HCN and HNC) undergo freeze-out as the number density of H$_{2}$ increases. However, current chemical models cannot explain other observed trends, such as the fact that the abundance of HCNH$^{+}$ shows an anti-correlation with that of HCN and HNC, but a positive correlation with that of N$_{2}$H$^{+}$ in the southern part of the Serpens South northern clump. This indicates that additional chemical pathways have to be invoked for the formation of HCNH$^{+}$ via molecules like N$_{2}$ in regions in which HCN and HNC freeze out.}
{Both the fact that HCNH$^{+}$ is most abundant in molecular cores prior to gravitational collapse and the fact that low-$J$ HCNH$^{+}$ transitions have very low H$_{2}$ critical densities make this molecular ion an excellent probe of pristine molecular gas.}


\titlerunning{Protonated hydrogen cyanide as a tracer of pristine molecular gas}

\authorrunning{Y. Gong et al.}

\maketitle

\section{Introduction}
Protonated hydrogen cyanide or iminomethylium, HCNH$^{+}$, is one of the fundamental molecular ions in the interstellar medium (ISM). Ions play an important role in interstellar chemistry as essential intermediaries in the ion-neutral reactions that dominate gas-phase chemistry in cold molecular clouds \citep[e.g.,][and references therein]{2013ChRv..113.8710A}. The chemistry of HCNH$^{+}$ in dense and cold regions has been extensively investigated. This ion is mainly produced by ion-neutral reactions \citep{1990ApJ...349..376T,1991A&A...247..487S,2000MNRAS.311..869H,2014MNRAS.443..398L,2017MNRAS.470.3194Q,2021A&A...651A..94F}, for instance:

\begin{equation}\label{f.form1}
\begin{split}
{\rm C^{+} + NH_{3}} &\to {\rm HCNH^{+} + H} \;
\end{split}
\end{equation}
\begin{equation}\label{f.form2}
\begin{split}
{\rm CH_{3}^{+} + N} &\to {\rm HCNH^{+} + H} \;
\end{split}
\end{equation}
\begin{equation}\label{f.form3}
\begin{split}
{\rm H_{3}^{+} + HCN/HNC} &\to {\rm HCNH^{+} + H_{2}} \;
\end{split}
\end{equation}
\begin{equation}\label{f.form4}
\begin{split}
{\rm HCO^{+} + HCN/HNC} &\to {\rm HCNH^{+} + CO} \;
\end{split}
\end{equation}
\begin{equation}\label{f.form5}
\begin{split}
{\rm H_{3}O^{+} + HCN/HNC} &\to {\rm HCNH^{+} + H_{2}O} \;
\end{split}
\end{equation}
\begin{equation}\label{f.form6}
\begin{split}
{\rm H_{2} + HCN^{+}/HNC^{+}} &\to {\rm HCNH^{+} + H} \;
\end{split}
\end{equation}
\begin{equation}\label{f.form7}
\begin{split}
{\rm C_{2}N^{+} + H_{2} } &\to {\rm HCNH^{+} + C} \;.
\end{split}
\end{equation}


On the other hand, HCNH$^{+}$ is thought to be mainly destroyed by dissociative recombination reactions:
\begin{equation}\label{f.des}
\begin{split}
{\rm HCNH^{+} + e^{-}} &\to {\rm HCN + H} \\
 &\to {\rm HNC + H} \\
 &\to {\rm CN + H + H}  \;.
\end{split}
\end{equation}
Reactions~(\ref{f.des}) are exothermic, which suggests that HCNH$^{+}$ is the main precursor of CN, HCN, and HNC \citep{1973ApJ...185..505H,1978ApJ...222..508H}. The chemical pathway of forming CN is energetically unfavorable in cold molecular clouds, because the simultaneous dissociation of the C-H and N-H bonds requires two electron excitations \citep[e.g.,][]{1998JChPh.108..698S}. Theoretical and observational studies support the notions that reactions~(\ref{f.des}) are the main pathways to produce HCN and HNC \citep[e.g.,][]{1998JChPh.108..698S,1998ApJ...503..717H,2002A&A...381..783A}, and that the branching ratios to produce HCN and HNC are nearly identical at low temperatures \citep[e.g.,][]{2012ApJ...746L...8M,2014MNRAS.443..398L,2021A&A...651A..94F}. 




HCNH$^{+}$ is a simple linear molecule with a small dipole moment of 0.29 D \citep{1986CPL...124..382B}. It was first discovered in Sgr B2 via the detection of its three lowest rotational transitions \citep{1986ApJ...302L..31Z}, and was subsequently observed in many other dense molecular clouds \citep[e.g.,][]{1991A&A...247..487S,1992ApJ...397L.123Z,2008ApJ...684.1221H,2017MNRAS.470.3194Q,2021A&A...651A..94F}. This suggests that HCNH$^{+}$ is ubiquitous in molecular clouds. 

Due to the nonzero nuclear spin of nitrogen, HCNH$^{+}$ transitions show electric quadrupole hyperfine structure (HFS) splitting. Spectra in which the HFS of the HCNH$^{+}$ $J = 1-0$ transition is resolved were presented, for example, by \citet{1992ApJ...397L.123Z} for TMC1\footnote{TMC1 harbors two chemically different dense cores which are known as the cyanopolyyne peak and the ammonia peak \citep[e.g.,][]{Toelle1981,1997ApJ...486..862P}. Throughout this paper, when we refer to TMC1, we mean to the cyanopolyyne peak, toward which the spectrum shown by \citet{1992ApJ...397L.123Z} was taken, rather than the ammonia peak.} and by \citet{2017MNRAS.470.3194Q} for L1544, which are both chemically well-studied dark clouds.
\citet{2021A&A...651A..94F} performed a survey of HCNH$^{+}$ (3--2) toward 26 high-mass star-forming cores, and reported the detection of this molecule in 16 targets. This study and the ones metioned above suggest that the typical fractional abundance of HCNH$^{+}$ with respect to H$_{2}$ spans a range of two orders of magnitude, from 10$^{-11}$ to 3$\times$10$^{-9}$. \citet{2017MNRAS.470.3194Q} found that the observed HCNH$^{+}$ and HC$_{3}$NH$^{+}$ abundances in the dark cloud L1544 could not be well reproduced at the same time by astrochemical models, indicating that the chemistry related to HCNH$^{+}$ is still not fully understood. Numerical simulations suggest that observations of the chemically co-evolving pair, HCN and HCNH$^{+}$, could be used to measure the ion-neutral drift velocity caused by ambipolar diffusion \citep{2023MNRAS.521.5087T}, which plays an important role in star formation \citep[e.g.,][]{1999ASIC..540..305M}. 

Measurements using the Cassini space probe indicate that HCNH$^{+}$ may also be the most abundant ion in the atmosphere of the largest satellite of Saturn, Titan \citep{2006GeoRL..33.7105C}. This molecular ion was not detected in Comet Hale-Bopp \citep[also known as C/1995 O1;][]{1999ApJ...527L..67Z}. These authors reported an upper limit of 1.9$\times 10^{13}$ cm$^{-2}$ for its column density and an abundance ratio [HCN]/[HCNH$^{+}$] of $\gtrsim$1 in this comet. 
Given all this, a thorough understanding of the chemistry of HCNH$^{+}$ in molecular clouds is still elusive, but is essential for us to gain insights into the process of chemical evolution of the ISM and star formation.


 Observations have shown that the environment of star-forming regions can strongly affect their chemistry \citep[e.g.,][]{2020ARA&A..58..727J}. However, the influence of the environment on the HCNH$^{+}$ fractional abundance is still unknown. 
 From an observational perspective, single-pointing studies of dense cores can result in loose constraints, whereas mapping studies can better address this issue. So far, HCNH$^{+}$ has rarely been mapped. Because HCNH$^{+}$ (3--2) is blended with an unidentified line at 222325 MHz (possibly from CH$_{3}$OCH$_{3}$) in the study of Sgr B2 in \citet{1991A&A...247..487S} due to the broad line widths, the map of the blended line can hardly be used to reliably trace the distribution of HCNH$^{+}$ in this cloud. Therefore, the spatial  distribution of HCNH$^{+}$ in molecular clouds remains largely unexplored.

At a distance of $\sim$440 pc \citep{2010ApJ...718..610D,2017ApJ...834..143O,2018ApJ...869L..33O,2019ApJ...879..125Z,2023A&A...673L...1O}, Serpens South and the Serpens filament are known as two of the nearest infrared dark clouds (IRCDs). Both only harbor low-mass young stellar objects \citep[YSOs; e.g.,][]{2008ApJ...673L.151G,2018A&A...620A..62G} and both contain parts that are quiescent as well as parts with active star formation \citep[e.g.,][]{2013MNRAS.436.1513F,2021A&A...646A.170G}, which allows us to probe the influence of star formation on the chemistry. 

Serpens South contains one of the nearest and youngest stellar protoclusters, the Serpens South cluster (SSC), while its northern clump (labeled as SSN hereafter) is quiescent. There are powerful outflows arising in SSC \citep[e.g.,][]{2011ApJ...737...56N,2014ApJ...791L..23N,2015ApJ...803...22P,2015Natur.527...70P,2015ApJS..219...21Z,2017A&A...600A..99M}, at the base of one of which H$_{2}$O masers are located \citep{2021AJ....162...68O}. 
In contrast, SSN is found to have high abundances of carbon-chain molecules \citep{2013MNRAS.436.1513F,2016ApJ...833..204F,2016ApJ...824..136L}; recently  even benzonitrile, c-C$_{6}$H$_{5}$CN, has been detected  \citep{2021NatAs...5..181B}. 

A similar trend is also observed in the Serpens filament that is quiescent in the southeast and shows signs of star formation in the northwest \citep{2018A&A...620A..62G,2021A&A...646A.170G}. Star formation in Serpens South is much more active than in the Serpens filament, which allows us to probe different levels of star formation feedback on the chemistry. Furthermore, the two regions have typical line widths of $\lesssim$2~\kms\,\citep[e.g.,][]{2013A&A...553A..58L,2014A&A...567A..78L,2016ApJ...833..204F,2021A&A...646A.170G}, which reduces the level of line confusion, which can be a significant problem for observations of the HCNH$^{+}$ (3--2) transition in regions like Sgr B2 \citep{1991A&A...247..487S}. Therefore, the Serpens filament and Serpens South are ideal sites to study the influence of the environment on the HCNH$^{+}$ fractional abundance. In this work, we present a mapping study of HCNH$^{+}$ to unveil its distribution and characterize its chemistry in these two regions.

Our observations are described in Sect.~\ref{Sec:obs}. In Sect.~\ref{Sec:res}, we report our results, which are discussed in Sect.~\ref{Sec:dis}. Lastly, our summary and conclusions are presented in Sect.~\ref{Sec:sum}.

\section{Observations and data reduction}\label{Sec:obs}
\subsection{IRAM-30 m and APEX-12 m observations}
We carried out HCNH$^{+}$ (1--0), HCNH$^{+}$ (2--1), H$^{13}$CN (1--0), HN$^{13}$C (1--0), H$^{13}$CO$^{+}$ (1--0), and N$_{2}$H$^{+}$ (1--0) imaging observations toward Serpens South and the Serpens filament with the IRAM-30m telescope\footnote{This work is based on observations carried out under project numbers 023-19, 127-20, 028-21 with the IRAM-30 m telescope. IRAM is supported by INSU/CNRS (France), MPG (Germany) and IGN (Spain).} during 2019 September 27--29, 2020 December 26--27, 2021 August 21--24, and October 22--25. During the observations, the EMIR dual-sideband and dual-polarization receivers (E090 and E150) were used as frontend \citep{2012A&A...538A..89C}, while Fast Fourier Transform Spectrometers (FFTSs) were used as backend. For the HCNH$^{+}$ (1--0) and HCNH$^{+}$ (2--1) lines, we used the narrow FFTSs, which provided a channel width of 50~kHz, corresponding to a velocity spacing of 0.2~\kms\,at 74~GHz and 0.1~\kms\,at 148~GHz (see also Table~\ref{Tab:lin}). For H$^{13}$CN (1--0), HN$^{13}$C (1--0), H$^{13}$CO$^{+}$ (1--0), and N$_{2}$H$^{+}$ (1--0), FFTSs with a channel width of 50~kHz and 250~kHz are used for the Serpens filament and Serpens South, respectively. The corresponding channel spacings in units of \kms\,are given in Table~\ref{Tab:lin}.


Using the Atacama Pathfinder EXperiment 12 meter  submillimeter telescope \citep[APEX\footnote{This publication is based on data acquired with the Atacama Pathfinder Experiment (APEX). APEX is a collaboration between the Max-Planck-Institut f{\"u}r Radioastronomie, the European Southern Observatory, and the Onsala Space Observatory.};][]{2006A&A...454L..13G}, we carried out HCNH$^{+}$ (3--2) mapping observations toward the Serpens filament during 2019 April 29 -- May 1 (project code: M9511A\_103) and pointed HCNH$^{+}$ (3--2) observations toward Serpens South during 2018 May 25 -- 2018 July 13 (project code: M9506A\_101). The coordinates of the pointed positions are given in Table~\ref{Tab:obspar}. A PI230 sideband separated and dual-polarization receiver\footnote{\url{https://www.eso.org/public/teles-instr/apex/pi230/}}, built by the Max Planck Institute for Radio Astronomy, was employed as frontend, while FFTSs were adopted as backend  \citep{2012A&A...542L...3K}. Each FFTS encompasses a bandwidth of 4~GHz and 65536 channels, resulting in a channel width of 61~kHz. The corresponding velocity spacing is 0.08~\kms\,for HCNH$^{+}$ (3--2) (see also Table~\ref{Tab:lin}). The observations were performed in the position-switching or On-The-Fly mode using the APECS software \citep{2006A&A...454L..25M}.

At the beginning of each observing session, pointing and focus were investigated toward bright continuum sources like Mars and Mercury. Pointing was regularly checked on nearby bright sources every hour, and its accuracy was found to be  within 5\arcsec. A single on-off observation toward SSC was performed to verify each frequency setup before mapping. The regions were observed in two orthogonal directions using the On-The-Fly mode. The standard chopper wheel method was used for calibrations and correcting the atmospheric attenuation of the IRAM-30 m observations \citep{1976ApJS...30..247U}, while an extension of the method using three loads was applied to the APEX-12 m observations. The calibration was undertaken about every 10 minutes. The antenna temperature, $T^{*}_{\rm A}$, was converted to main beam temperature, $T_{\rm mb}$, by applying the relation, $T_{\rm mb}=T^{*}_{\rm A}\eta_{\rm f}/\eta_{\rm mb}$, where $\eta_{\rm f}$ and $\eta_{\rm mb}$ are the forward efficiency and main beam efficiency\footnote{see \url{https://publicwiki.iram.es/Iram30mEfficiencies} and \url{http://www.apex-telescope.org/telescope/efficiency/}}, respectively (see Table~\ref{Tab:lin} for the values). The uncertainties in the absolute flux calibration are assumed to be 10\%\,in this work. The observed spectral lines, their rest frequencies, and other spectroscopic and observational information 
are listed in Table~\ref{Tab:lin}. Velocities with respect to the local standard of rest (LSR) are reported throughout this work.  

The GILDAS\footnote{\url{https://www.iram.fr/IRAMFR/GILDAS/}} software was used to reduce the spectral line data \citep{2005sf2a.conf..721P}, and a first-order baseline was subtracted from each spectrum. For all the maps, raw spectra were gridded into a data cube with a pixel size of 5\arcsec$\times$5\arcsec. Since  the HCNH$^{+}$ emission in our sources is generally weak and has a smooth extent, all spectral data are smoothed to a half power beam width (HPBW) of 36$\rlap{.}$\arcsec3. This facilitates the comparison with the \textit{Herschel} H$_{2}$ column density and dust temperature maps (see Sect.~\ref{Sec:arc}).

\begin{table*}[!hbt]
\caption{Observational and spectroscopic parameters related to the molecular lines presented in this work.}\label{Tab:lin}
\tiny
\centering
\begin{tabular}{ccccccccccccc}
\hline \hline
             &         &      &   &    &   &  & & \multicolumn{2}{c}{Serpens filament}  &  \multicolumn{2}{c}{Serpens South}   &    \\ 
\cmidrule(lr){9-10} 
\cmidrule(lr){11-12}
Transition             & Frequency         & $E_{\rm u}/k$  & $A$ & $\mu$ & $n_{\rm c}({\rm H}_{2})$ & $\theta_{\rm beam}$  & $\eta_{\rm mb}/\eta_{\rm f}$   & $\sigma$ & $\delta \varv$  & $\sigma$ & $\delta \varv$   &  Telescope  \\ 
                 & (MHz)             & (K)      & (s$^{-1}$)    & (D)  & (cm$^{-3}$) &  (\arcsec)       &    & (mK) & (\kms)                  &  (mK) & (\kms)  &         \\ 
(1)              & (2)               & (3)              & (4)              & (5) & (6)  & (7)                     & (8)     & (9) & (10) &  (11)  & (12) & (13)\\
\hline
HCNH$^{+}$ $J=1-0$ & 74111.302(3)         & 3.6  & $1.4\times 10^{-7}$ & 0.29 & $2.0\times 10^{2}$ & 33 & 0.87 & -- & -- & 51 & 0.20  & IRAM-30 m        \\
HCNH$^{+}$ $J=2-1$ & 148221.450(17)       & 10.7 & $1.3\times 10^{-6}$ & 0.29 & $1.7\times 10^{3}$ & 18 & 0.78 & 21 & 0.21 & 58 & 0.21  & IRAM-30 m        \\   
HCNH$^{+}$ $J=3-2$ & 222329.277(8)        & 21.3 & $4.9\times 10^{-6}$ & 0.29 & $5.2\times 10^{3}$ & 27 & 0.80 & 28 & 0.08 & 20 & 0.08  & APEX             \\
H$^{13}$CN $J=1-0$ & 86339.921(1)         & 4.1  & $2.2\times 10^{-5}$ & 2.99 & $5.3\times 10^{5}$ & 28 & 0.85 & 22 & 0.17 & 24 & 0.68 & IRAM-30 m        \\
HN$^{13}$C $J=1-0$ & 87090.825(4)         & 4.2  & $1.9\times 10^{-5}$ & 2.70 & $1.4\times 10^{5}$ & 28 & 0.85 & 18 & 0.17 & 26 & 0.67 & IRAM-30 m        \\
H$^{13}$CO$^{+}$ $J=1-0$ & 86754.288(5)   & 4.2  & $3.9\times 10^{-5}$ & 3.90 & $6.2\times 10^{4}$& 28 & 0.85 & 40 & 0.17 & 27 & 0.68 & IRAM-30 m        \\
N$_{2}$H$^{+}$ $J=1-0$ & 93173.398(1)     & 4.5  & $3.6\times 10^{-5}$ & 3.40 & $6.1\times 10^{4}$& 26 & 0.84 & 19 & 0.16 & 26 & 0.63 & IRAM-30 m        \\
\hline
\end{tabular}
\tablefoot{(1) Observed transition. All transitions were observed in the On-The-Fly mode, except for HCNH$^{+}$ (3--2) which was observed as single pointing toward Serpens South. (2) Rest frequency taken from the Cologne Database for Molecular Spectroscopy \citep[CDMS\footnote{https://zeus.ph1.uni-koeln.de/cdms},][]{2016JMoSp.327...95E}. Uncertainties in the last digits are given in parentheses. For the observed lines, only the frequency of the strongest HFS component is listed. (3) Upper energy level of the observed transition. (4) Einstein A coefficient. (5) Dipole moment from CDMS \citep{2016JMoSp.327...95E}. (6) Optically thin H$_{2}$ critical density at a kinetic temperature of 10~K. The critical densities of HCNH$^{+}$ transitions are calculated using the collisional rate coefficients from \citet{2023JChPh.158g4304B} and Eq.~(4) in \citet{2015PASP..127..299S}. For the other lines, the values are directly taken from Table~1 in \citet{2015PASP..127..299S}. The critical density of HN$^{13}$C (1--0) is assumed to be the same as that of HNC (1--0). (7) HPBW. (8) $\eta_{\rm f}$ and $\eta_{\rm mb}$ are the forward efficiency and main beam efficiency, respectively. (9) RMS level at the smoothed angular resolution of 36\rlap{.}\arcsec3. (10) Velocity channel spacing. (11) RMS level at the smoothed angular resolution of 36\rlap{.}\arcsec3. For pointed HCNH$^{+}$ (3--2) observations in Serpens South, the RMS level is estimated at the original angular resolution of 27\arcsec. (12) Velocity channel spacing. (13) Telescope.}
\normalsize
\end{table*}

\subsection{Archival data}\label{Sec:arc}
Dust-based H$_{2}$ column density and dust temperature maps\footnote{\url{http://www.herschel.fr/cea/gouldbelt/en/Phocea/Vie_des_labos/Ast/ast_visu.php?id_ast=66}} were derived from spectral energy distributions (SEDs) fitted to \textit{Herschel} data at four far-infrared bands, that is, 160, 250, 350, and 500~$\mu$m \citep{2010A&A...518L.102A,2015A&A...584A..91K,2021MNRAS.500.4257F}. These maps also have an angular resolution of 36$\rlap{.}$\arcsec3. 

NH$_{3}$ and CO archival data are only available for Serpens South. 
For NH$_{3}$ in Serpens South, we utilized the data products\footnote{\url{https://dataverse.harvard.edu/dataset.xhtml?persistentId=doi:10.7910/DVN/SDHQRP}} from observations with the Green Bank Telescope \citep{2016ApJ...833..204F}. The corresponding HPBW is 32\arcsec. 
CO (3--2) data\footnote{\url{http://th.nao.ac.jp/MEMBER/nakamrfm/sflegacy/data.html}} of Serpens South were obtained with the Atacama Submillimeter Telescope Experiment (ASTE) 10~m telescope. The data that we retrieved has a HPBW of 22\arcsec\, and a channel width of $\sim$0.11~\kms. The typical noise level is 0.11~K on a $T_{\rm A}^{*}$ scale at a velocity spacing of 0.5~\kms. The main beam efficiency is 0.57 at 345 GHz. Details of the observations are presented in \citet{2011ApJ...737...56N}. 


\section{Results}\label{Sec:res}
\subsection{Molecular distribution}

\begin{figure*}[!htbp]
\centering
\includegraphics[width = 1 \textwidth]{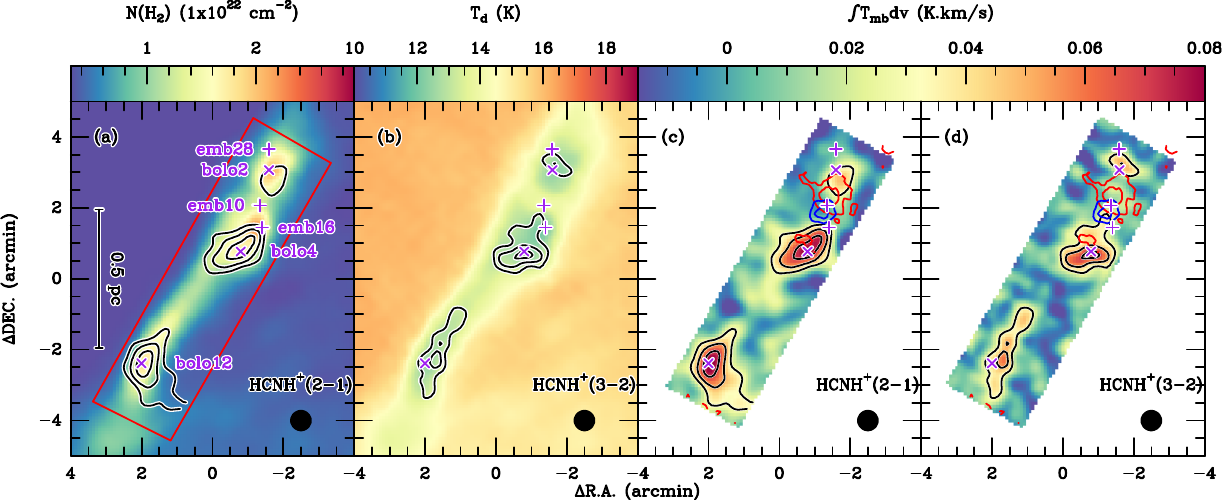}
\caption{{(a) \textit{Herschel} H$_{2}$ column density map of the Serpens filament \citep{2021MNRAS.500.4257F} overlaid with the HCNH$^{+}$ (2--1) integrated intensity contours. The velocity range over which the HCNH$^{+}$ (2--1) intensity map is integrated is 7.7--8.7~\kms. The contours start at 0.03~K~\kms\,(3$\sigma$) and increase by 0.02~K~\kms. The red boundary represents the region mapped by our IRAM 30-m and APEX 12-m observations. (b) \textit{Herschel} dust temperature map \citep{2021MNRAS.500.4257F} overlaid with the HCNH$^{+}$ (3--2) integrated intensity contours. The velocity range over which the HCNH$^{+}$ (3--2) intensity is integrated is 7.7--8.7~\kms. The contours start at 0.03~K~\kms\,(3$\sigma$) and increase by 0.02~K~\kms. (c) HCNH$^{+}$ (2--1) integrated intensity map overlaid with the blue-, and red-shifted outflow lobes of $^{13}$CO (2--1) that are driven by emb10 \citep{2021A&A...646A.170G}. The black contours are the same as in panel a. (d) Similar to panel c, but for HCNH$^{+}$ (3--2). The beam size is shown in the lower right corner of each panel. In all panels, the (0, 0) offset corresponds to $\alpha_{\rm J2000}$=18$^{\rm h}$28$^{\rm m}$50$\rlap{.}^{\rm s}$4, $\delta_{\rm J2000}$=00$^{\circ}$49$^{\prime}$58$\rlap{.}^{\prime \prime}$72. The three purple plus signs indicate the positions of the three embedded YSOs, emb10, emb16, and emb 28 \citep{2009ApJ...692..973E}, while the three purple crosses mark the positions of three 1.1~mm dust continuum cores, bolo2, bolo4, and bolo12 \citep{2007ApJ...666..982E}.}\label{Fig:sf-hcnh+}}
\end{figure*}

Figure~\ref{Fig:sf-hcnh+} presents the distributions of HCNH$^{+}$ (2--1) and HCNH$^{+}$ (3--2) emission toward the Serpens filament. The overall distributions of the two transitions are similar in the Serpens filament, with slight differences arising from the fact that the signal-to-noise ratio of HCNH$^{+}$ (3--2) is lower than that of HCNH$^{+}$ (2--1). HCNH$^{+}$ is detected toward three dense cores, called bolo2, bolo4, and bolo12 by \citet{2007ApJ...666..982E}. The name ``bolo" refers to Bolocam that is a large-format bolometric camera of the Caltech Submillimeter Observatory \citep{2003SPIE.4855...30G}. The emission is the brightest in bolo4 followed by bolo12. Based on the comparison with the \textit{Herschel} results, we find that the detected HCNH$^{+}$ emission arises from regions with dust temperatures, $T_{\rm d}$, of $\lesssim$12~K and H$_{2}$ column densities $>$1$\times 10^{22}$~cm$^{-2}$ \citep{2021MNRAS.500.4257F}, which indicates that the observed HCNH$^{+}$ emission mainly arises from cold and high H$_{2}$ column density regions. Toward the YSO emb10\footnote{The term ``emb" is the abbreviation of ``embedded" \citep{2009ApJ...692..973E}.}, which drives a molecular outflow \citep[][see also  Fig.~\ref{Fig:sf-hcnh+}c and \ref{Fig:sf-hcnh+}d for the outflow lobes]{2021A&A...646A.170G,2023A&A...672C...1G}, the HCNH$^{+}$ emission appears to be absent. 

\begin{figure*}[!htbp]
\centering
\includegraphics[width = 1 \textwidth]{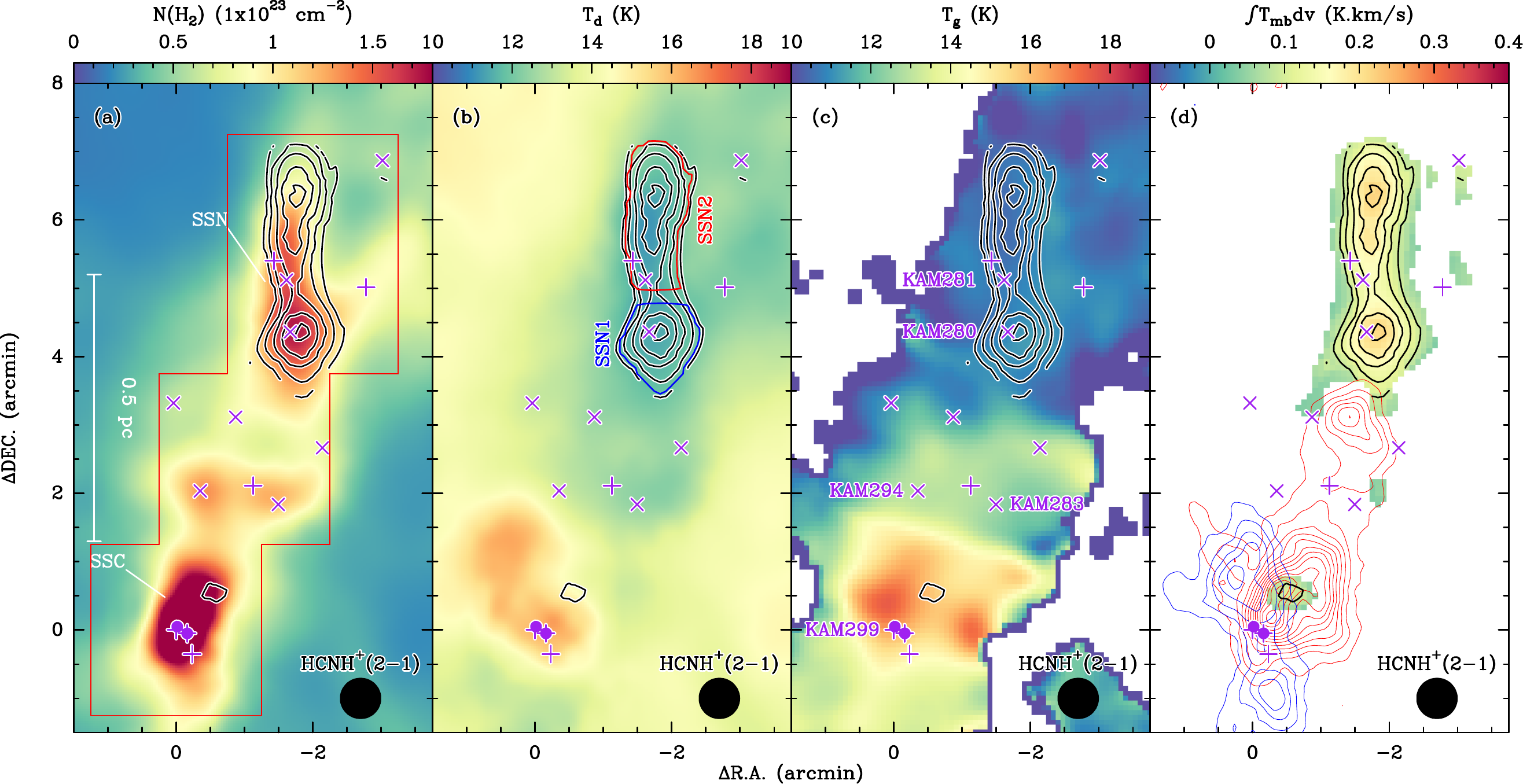}
\caption{{(a) \textit{Herschel} H$_{2}$ column density map of Serpens South \citep{2015A&A...584A..91K} overlaid with  HCNH$^{+}$ (2--1) integrated intensity contours. The velocity range over which the HCNH$^{+}$ (2--1) intensity is integrated is 6.5-8.5~\kms. The contours start at 0.09~K~\kms\,(3$\sigma$) and increase by 0.03~K~\kms. The red boundary represents the region mapped by our IRAM-30~m observations. (b) \textit{Herschel} dust temperature map \citep{2015A&A...584A..91K} overlaid with the HCNH$^{+}$ (2--1) integrated intensity contours that are the same as in panel a. SSN is further divided into two subregions, SSN1 and SSN2, which are labeled in this panel. (c) Gas kinetic temperature map derived from ammonia observations \citep{2016ApJ...833..204F} overlaid with the HCNH$^{+}$ (2--1) integrated intensity contours that are the same as in panel a. (d) HCNH$^{+}$ (2--1) integrated intensity map overlaid with the CO (3--2) blueshifted and redshifted outflow lobes from the SSC \citep{2011ApJ...737...56N}. The black contours represent HCNH$^{+}$ (2--1) integrated intensity as the contours in panel a. The beam size is shown in the lower right corner of each panel. In all panels, the (0, 0) offset corresponds to $\alpha_{\rm J2000}$=18$^{\rm h}$30$^{\rm m}$04$\rlap{.}^{\rm s}$19, $\delta_{\rm J2000}$=$-$02$^{\circ}$03$^{\prime}$05$\rlap{.}^{\prime \prime}$5. The two purple circles represent the two 22~GHz water masers \citep{2021AJ....162...68O}. The purple pluses indicate the positions of deeply embedded YSOs \citep{2015A&A...584A..91K,2022MNRAS.516.5244S}, while the purple crosses mark the positions of the prestellar dense cores extracted from the \textit{Herschel} dust continuum maps \citep{2015A&A...584A..91K}.}\label{Fig:sers-hcnhp}}
\end{figure*}

Figure~\ref{Fig:sers-hcnhp} presents the distribution of HCNH$^{+}$ (2--1) emission toward Serpens South with the purple crosses marking the positions of the prestellar dense cores extracted from the \textit{Herschel} dust continuum maps \citep{2015A&A...584A..91K}. We find that the HCNH$^{+}$ emission is prominent toward SSN, but barely detected toward SSC. The brightest HCNH$^{+}$ emission lies in the southern part of SSN (labeled as SSN1 in Fig.~\ref{Fig:sers-hcnhp}b), which is slightly offset from the dust continuum peak (the prestellar core KAM280, No. 280 in \citealt{2015A&A...584A..91K}, also named as SerS MM9 by \citealt{2011A&A...535A..77M}). The emission in the northern part of SSN (labeled as SSN2 in Fig.~\ref{Fig:sers-hcnhp}b) and the weak emission in SSC appear not to be associated with any dust continuum core. The detected HCNH$^{+}$ emission in SSN comes from regions with $T_{\rm d}\lesssim$12~K, similar to those in the Serpens filament. We also note that the weak HCNH$^{+}$ emission in SSC arises from a slightly warmer region with $T_{\rm d}\sim$15~K. The ammonia observations \citep{2016ApJ...833..204F} confirm that the gas kinetic temperatures are $\lesssim$12~K in SSN and $\sim$15~K in SSC. On the other hand, the HCNH$^{+}$ emitting region in Serpens South has H$_{2}$ column densities $\gtrsim$4$\times 10^{22}$~cm$^{-2}$. In Fig.~\ref{Fig:sers-hcnhp}d, the weak HCNH$^{+}$ emission in SSC appears to coincide with the peak of the redshifted outflow lobe \citep{2011ApJ...737...56N} that is about 30\arcsec\,offset from the H$_{2}$ column density peak. However, the observed line width of the weak emission is as narrow as 0.45$\pm$0.21~\kms, which does not suggest that the weak HCNH$^{+}$ emission is produced by the outflow shocks. 

\begin{figure*}[!htbp]
\centering
\includegraphics[width = 0.9 \textwidth]{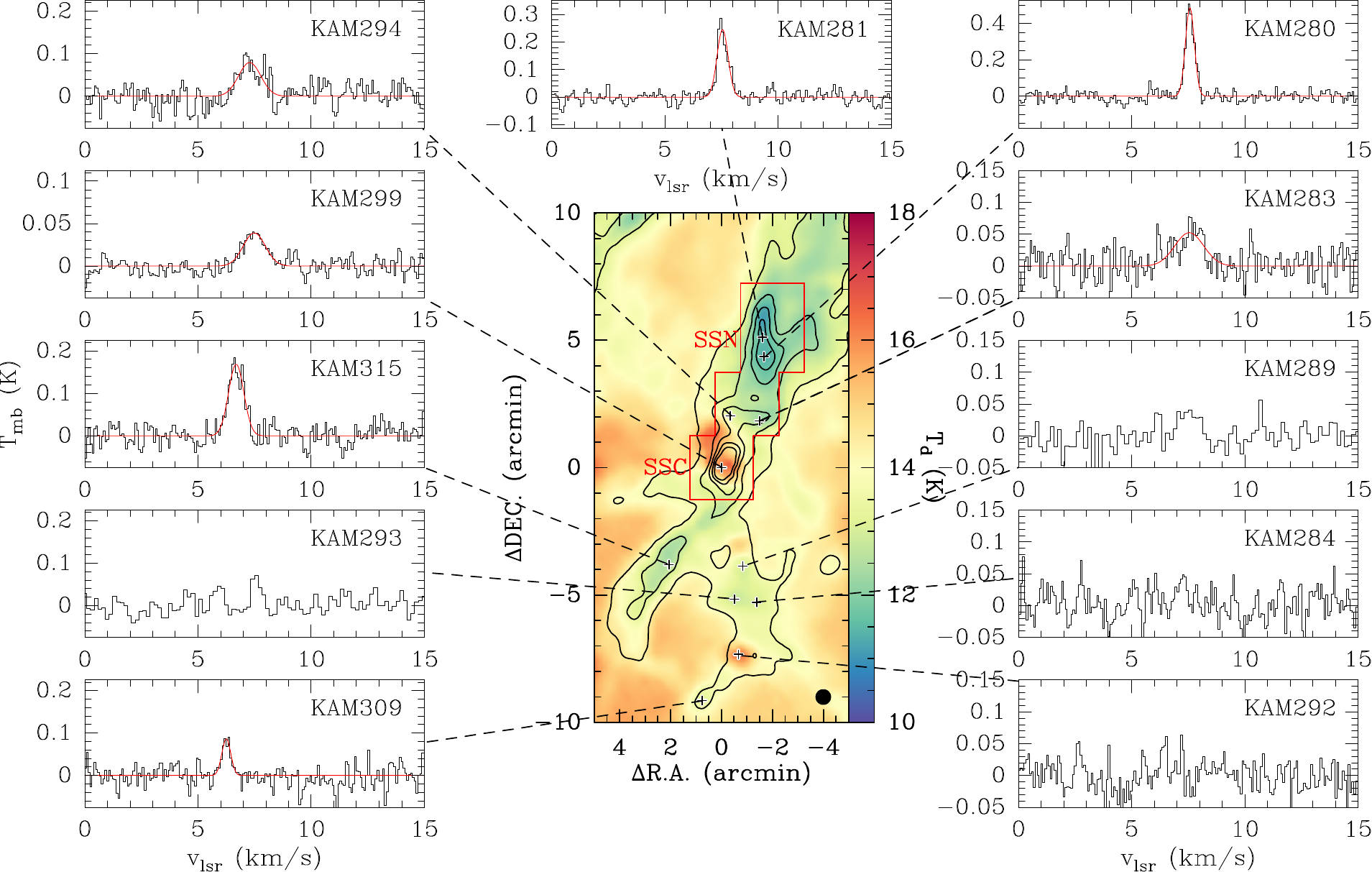}
\caption{{\textit{Herschel} dust temperature map of Serpens South \citep{2015A&A...584A..91K} overlaid with H$_{2}$ column density contours that start at 2$\times 10^{22}$ cm$^{-2}$ and increase by 2$\times 10^{22}$ cm$^{-2}$. The beam size is shown in the lower right corner of the central panel. The (0, 0) offset corresponds to $\alpha_{\rm J2000}$=18$^{\rm h}$30$^{\rm m}$04$\rlap{.}^{\rm s}$19, $\delta_{\rm J2000}$=$-$02$^{\circ}$03$^{\prime}$05$\rlap{.}^{\prime \prime}$5. The region mapped in HCNH$^{+}$ (2--1) is indicated by the red boundary. The positions toward which pointed observations of HCNH$^{+}$ (3--2) were carried out are indicated by the black plus signs in Serpens South. To the spectra: observed HCNH$^{+}$ (3--2) spectra are displayed in black, 
while Gaussian fits to the line profiles are shown in red. The source name is indicated in the upper right corner of each panel.}\label{Fig:ss-hcnh+}}
\end{figure*}

As shown in Fig.~\ref{Fig:ss-hcnh+}, our APEX-12 m pointed observations toward Serpens South have led to the detection of HCNH$^{+}$ (3--2) in seven dense cores with $T_{\rm d}\lesssim$18~K. The brightest HCNH$^{+}$ (3--2) emission comes from KAM280, which is in agreement with our HCNH$^{+}$ (2--1) map. The detection of HCNH$^{+}$ (3--2) in KAM299 (also known as SerS-MM18) suggests that HCNH$^{+}$ exists in SSC, but that its emission is much weaker than in SSN. The observed line profiles are commonly Gaussian-like. In order to derive the observational parameters, we employ a single-Gaussian component to fit the observed spectra, because the HFS lines of the 3--2 transition are too close in frequency to be separated. The fitted peak main beam brightness temperature, velocity centroids, and full width at half maximum (FWHM) line widths are given in Table~\ref{Tab:obspar}.

\begin{figure}[!htbp]
\centering
\includegraphics[width = 0.45 \textwidth]{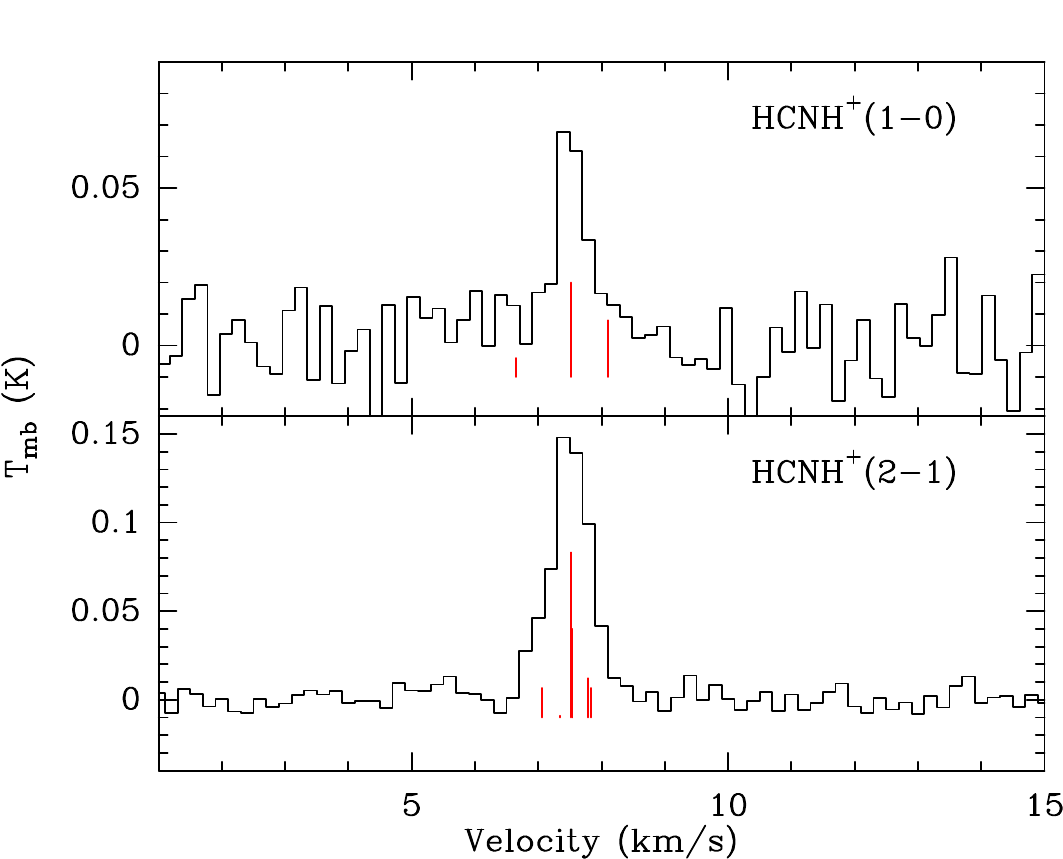}
\caption{{HCNH$^{+}$ (1--0) and HCNH$^{+}$ (2--1) spectra averaged over the region where HCNH$^{+}$ (2--1) integrated intensities are higher than 3$\sigma$ in Fig.~\ref{Fig:sers-hcnhp}. The positions and relative intensities of the HFS components are indicated by the red vertical lines.}\label{Fig:avsp}}
\end{figure}

We also mapped Serpens South in HCNH$^{+}$ (1--0) but no emission was detected. This places a 3$\sigma$ upper limit of 0.15~K for the peak intensity. In order to detect the weak HCNH$^{+}$ (1--0) emission, we average all the HCNH$^{+}$ (1--0) spectra in the region where the HCNH$^{+}$ (2--1) integrated intensities are higher than 3$\sigma$. The results are shown in Fig.~\ref{Fig:avsp}. The HCNH$^{+}$ (1--0) average spectrum has a $\sim$5$\sigma$ detection and peaks at exactly the same velocity as its HCNH$^{+}$ (2--1) counterpart, securing the detection of HCNH$^{+}$ (1--0). However, the contributions of the HFS satellite lines are not discernible in Fig.~\ref{Fig:avsp} at the current noise level. The detected lines are mainly attributed to the contributions of the $F=2-1$ and $F=3-2$ hyperfine lines for $J=1-0$ and $J=2-1$, respectively. 


\begin{table*}[!hbt]
\caption{Spectral line properties of the HCNH$^{+}$ (3--2) transition observed toward the selected positions in Serpens South.}\label{Tab:obspar}
\normalsize
\centering
\begin{tabular}{cccccccc}
\hline \hline
Name             & $\alpha_{\rm J2000}$ & $\delta_{\rm J2000}$  &  $T_{\rm p}$ &  $\varv_{\rm lsr}$   & $\Delta \varv$ & Class  & Other names \\ 
                 & (h~m~s)             & (\degree~\arcmin~\arcsec)              &  (mK)       &  (\kms)  & (\kms) &              &      \\ 
(1)              & (2)               & (3)              & (4)              & (5)    & (6)   &    (7)        & (8)         \\
\hline
KAM280        & 18:29:57.50 & $-$01:58:43.7 & 485$\pm$33 & 7.57$\pm$0.01  & 0.47$\pm$0.02 & prestellar & SerS-MM9  \\
KAM281        & 18:29:57.72 & $-$01:57:58.2 & 245$\pm$30 & 7.55$\pm$0.01  & 0.58$\pm$0.03 & prestellar & SerpS-MM8 \\
KAM283        & 18:29:58.20 & $-$02:01:15.2 & 52$\pm$18  & 7.55$\pm$0.01  & 1.49$\pm$0.34 & prestellar & \\
KAM284        & 18:29:58.66 & $-$02:08:22.5 & $<$67      & --             & -- & prestellar & \\
KAM289        & 18:30:00.85 & $-$02:06:57.3 & $<$56      & --             & -- & protostellar & SerpS-MM11 \\
KAM292        & 18:30:01.50 & $-$02:10:25.5 & $<$56      & --             & -- & protostellar & SerpS-MM13, IRAS 18274-0212\\
KAM293        & 18:30:02.13 & $-$02:08:15.1 & $<$55      & --             & -- & prestellar & SerS-MM14 \\
KAM294        & 18:30:02.77 & $-$02:01:03.4 & 79$\pm$25  & 7.28$\pm$0.06  & 1.15$\pm$0.12 & prestellar &	SerS-MM15\\
KAM299        & 18:30:04.19 & $-$02:03:05.5 & 39$\pm$5   & 7.49$\pm$0.05  & 1.16$\pm$0.11 & protostellar & SERS~02, SerpS-MM18\\
KAM309        & 18:30:07.24 & $-$02:12:13.6 & 84$\pm$10  & 6.25$\pm$0.04  & 0.44$\pm$0.09 & prestellar & \\
KAM315        & 18:30:12.44 & $-$02:06:53.6 & 169$\pm$21 & 6.69$\pm$0.02  & 0.80$\pm$0.05 & protostellar & SerpS-MM22 \\
\hline
\end{tabular}
\tablefoot{(1) Source name from \citet{2015A&A...584A..91K}. (2) Right ascension. (3) Declination. (4) Peak main beam brightness temperature. (5) Velocity centroid. (6) FWHM line width. (7) Classification from \citet{2015A&A...584A..91K}. (8) Other names from \citet{2011A&A...535A..77M} and \citet{2017A&A...600A..99M}.}
\normalsize
\end{table*}

\begin{figure*}[!htbp]
\centering
\includegraphics[width = 1 \textwidth]{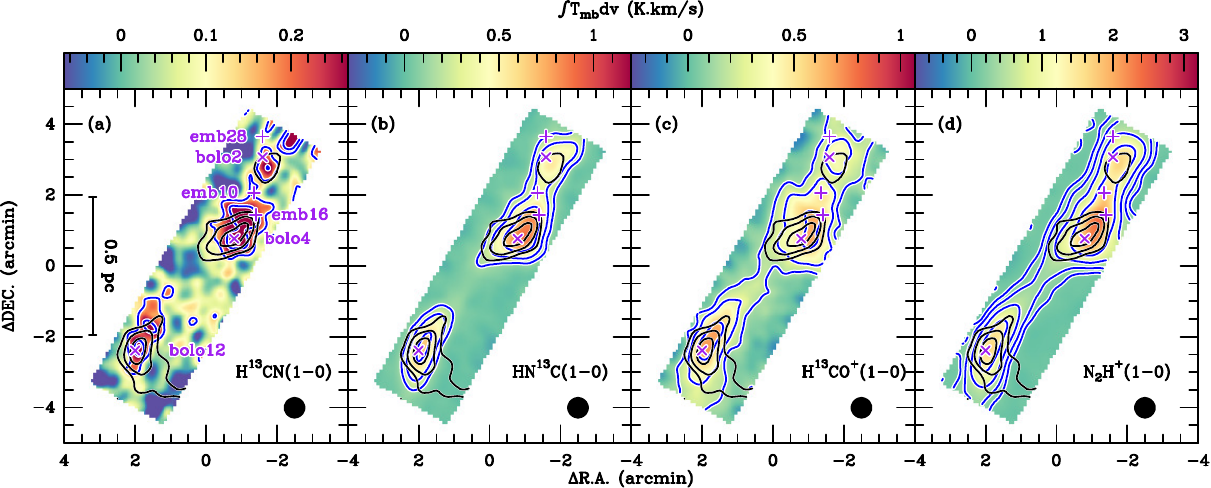}
\caption{{Comparison of different tracers in the Serpens filament. Integrated-intensity maps (in color scale) of (a) H$^{13}$CN (1--0), (b) HN$^{13}$C (1--0), (c) H$^{13}$CO$^{+}$(1--0), and (d) N$_{2}$H$^{+}$ (1--0). The blue contours represent the respective integrated intensities that are $2^{n}\times $0.15~K~\kms and $n=$0, 1, 2,... The black contours of HCNH$^{+}$ (2--1) integrated intensities are the same as in  Fig.~\ref{Fig:sf-hcnh+}a. The intensity was integrated over 7--9.5~\kms for HN$^{13}$C (1--0), and H$^{13}$CO$^{+}$(1--0), while a broader velocity range (5--9.5~\kms) was chosen to cover the three main HFS components of N$_{2}$H$^{+}$ (1--0). The beam size is shown in the lower right corner of each panel. In all panels, the (0, 0) offset corresponds to $\alpha_{\rm J2000}$=18$^{\rm h}$28$^{\rm m}$50$\rlap{.}^{\rm s}$4, $\delta_{\rm J2000}$=00$^{\circ}$49$^{\prime}$58$\rlap{.}^{\prime \prime}$72. The markers are the same as in  Fig.~\ref{Fig:sf-hcnh+}.}\label{Fig:sf-tracer}}
\end{figure*}

\begin{figure*}[!htbp]
\centering
\includegraphics[width = 1 \textwidth]{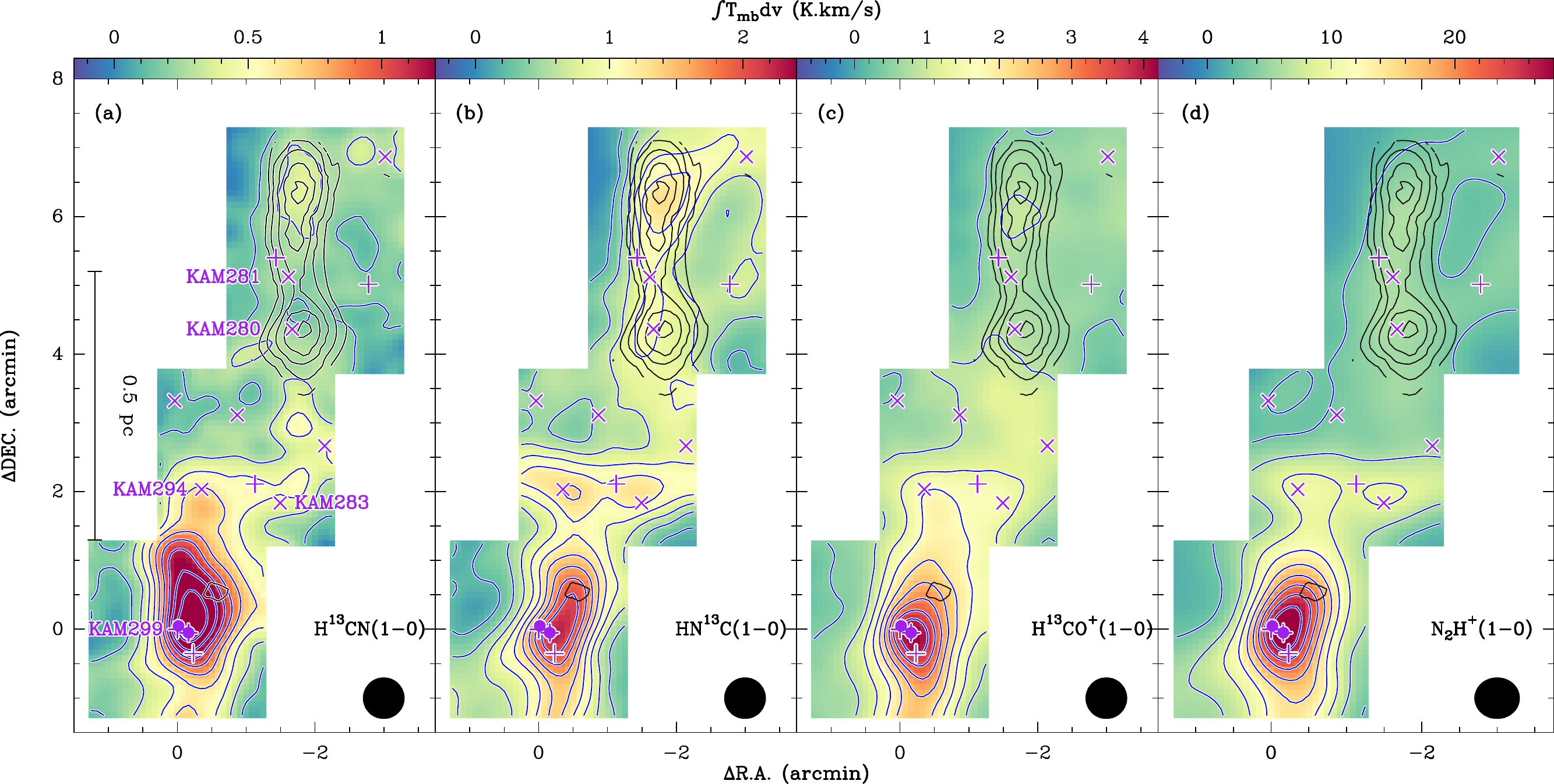}
\caption{{Comparison of different tracers in Serpens South. Integrated-intensity maps (in color scale) of (a) H$^{13}$CN (1--0), (b) HN$^{13}$C (1--0), (c) H$^{13}$CO$^{+}$(1--0), and (d) N$_{2}$H$^{+}$ (1--0). The blue contours represent the respective integrated intensities. They vary from 10\% to 90\% of the peak integrated intensities in steps of 10\%, where the peak integrated intensities are 1.57~K~\kms, 2.25~K~\kms, 4.26~K~\kms, and 30.24~K~\kms\,for H$^{13}$CN (1--0), HN$^{13}$C (1--0),  H$^{13}$CO$^{+}$(1--0), and N$_{2}$H$^{+}$ (1--0), respectively.
The black contours of HCNH$^{+}$ (2--1) integrated intensities are the same as in Fig.~\ref{Fig:sers-hcnhp}a.
The intensity weas integrated over 7--9.5~\kms for H$^{13}$CN (1--0), HN$^{13}$C (1--0), and H$^{13}$CO$^{+}$(1--0), while a broader range (5--9.5~\kms) was chosen to cover the three main HFS components of N$_{2}$H$^{+}$ (1--0). The beam size is shown in the lower right corner of each panel. In all panels, the (0, 0) offset corresponds to $\alpha_{\rm J2000}$=18$^{\rm h}$30$^{\rm m}$04$\rlap{.}^{\rm s}$19, $\delta_{\rm J2000}$=$-$02$^{\circ}$03$^{\prime}$05$\rlap{.}^{\prime \prime}$5. The markers are the same as in  Fig.~\ref{Fig:sers-hcnhp}.}\label{Fig:sers-tracer}}
\end{figure*}

In this work, we use H$^{13}$CN (1--0), HN$^{13}$C (1--0), H$^{13}$CO$^{+}$ (1--0), and  N$_{2}$H$^{+}$ (1--0) emission as proxies to investigate the distributions of HCN, HNC, HCO$^{+}$, and N$_{2}$H$^{+}$ emission toward the Serpens filament and Serpens South. The Serpens filament has been mapped in N$_{2}$H$^{+}$ (1--0) by \citet{2018A&A...620A..62G}. Serpens South has been mapped in N$_{2}$H$^{+}$ (1--0) and H$^{13}$CO$^{+}$ (1--0) by \citet{2013ApJ...766..115K} and \citet{2013ApJ...778...34T}. In comparison to these studies, the observations presented here provide either higher angular resolutions or better sensitivities or both. 
Figures~\ref{Fig:sf-tracer}--\ref{Fig:sers-tracer} present the integrated intensity maps of HCNH$^{+}$, H$^{13}$CN, HN$^{13}$C, H$^{13}$CO$^{+}$, and N$_{2}$H$^{+}$. Comparing the distribution of HCNH$^{+}$ with those of the other tracers in the Serpens filament (see Fig.~\ref{Fig:sf-tracer}), we find that the different tracers have nearly the same distribution toward bolo12, which is not affected by star formation. Toward bolo4, the distribution of HCNH$^{+}$ is similar to those of H$^{13}$CN and HN$^{13}$C, but is slightly different from those of H$^{13}$CO$^{+}$ and N$_{2}$H$^{+}$ (i.e., the peaks of the H$^{13}$CO$^{+}$ and N$_{2}$H$^{+}$ emission lie to the west of the HCNH$^{+}$ emission peak). Since bolo4 lies close to the blueshifted outflow lobe from emb10 (see Fig.~\ref{Fig:sf-hcnh+}), the difference could be caused by outflow feedback. 

\begin{figure*}[!htbp]
\centering
\includegraphics[width = 1 \textwidth]{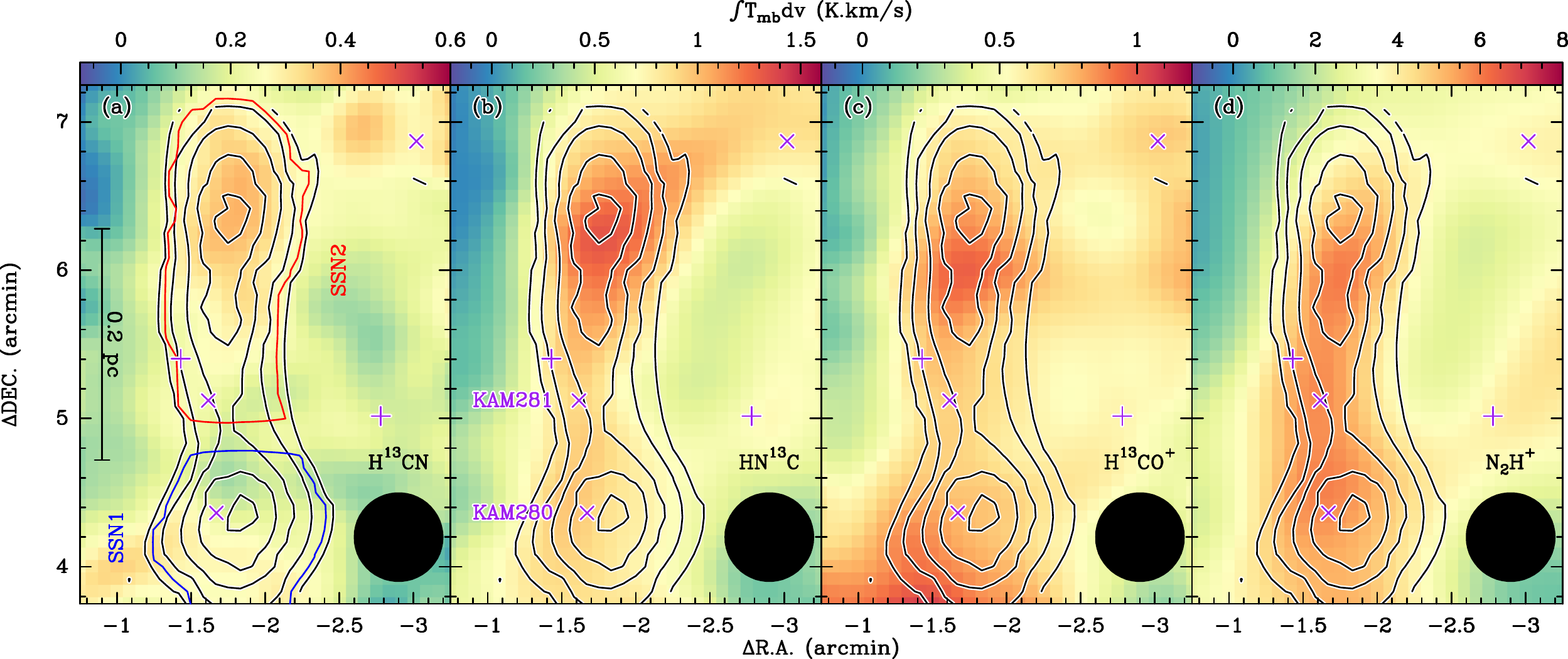}
\caption{{Same as Fig.~\ref{Fig:sers-tracer} but zoomed in on SSN with different color scales for better visualization. The two subregions, SSN1 and SSN2, are labeled in panel (a).}\label{Fig:sers-zoom}}
\end{figure*}

In Serpens South (Fig.~\ref{Fig:sers-tracer}) HCNH$^{+}$ emission is weak toward SSC, which is in stark contrast to the other four tracers which exhibit the brightest emission toward SSC. Toward SSN, the molecular morphologies are different for the various molecular gas tracers, which is more evident in Fig.~\ref{Fig:sers-zoom}.
No molecular line emission peaks at the positions of the dust peaks. The HCNH$^{+}$ emission peaks are offset from the emission peaks of other molecular line tracers, which could be (at least partially) caused by selective freeze-out processes and optical depth effects, as discussed further in Sect.~\ref{sec.col}. 



\subsection{Molecular column densities and abundances}\label{sec.col}
\begin{figure*}[!htbp]
\centering
\includegraphics[width = 1 \textwidth]{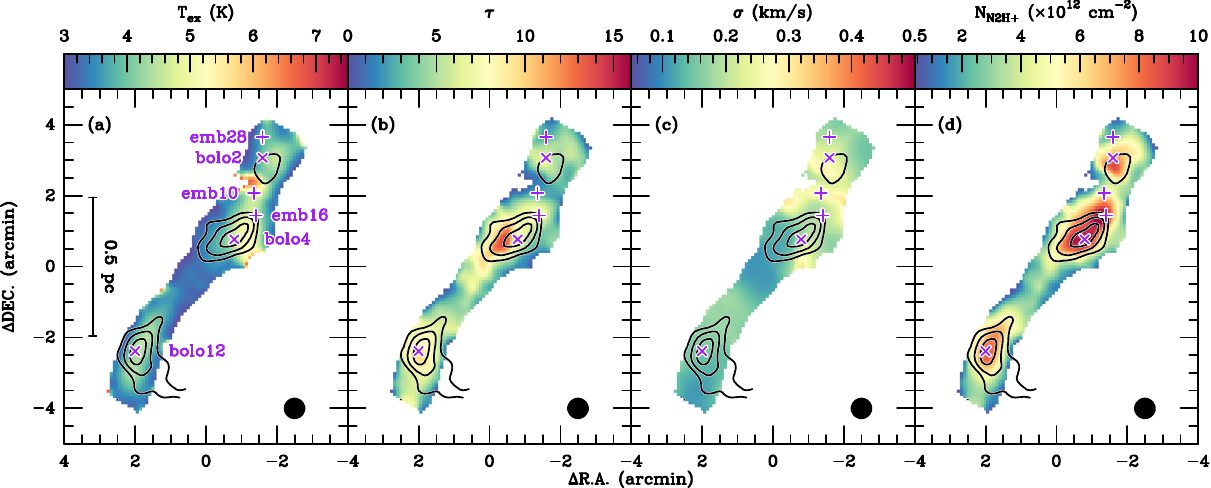}
\caption{{Distributions (a) N$_{2}$H$^{+}$ (1--0) excitation temperature, (b) total optical depth, (c) velocity dispersion, and (d) N$_{2}$H$^{+}$ column density toward the Serpens filament. The overlaid HCNH$^{+}$ (2--1) integrated-intensity contours are the same as in Fig.~\ref{Fig:sf-hcnh+}a.}\label{Fig:n2hp-serf}}
\end{figure*}

\begin{figure*}[!htbp]
\centering
\includegraphics[width = 1 \textwidth]{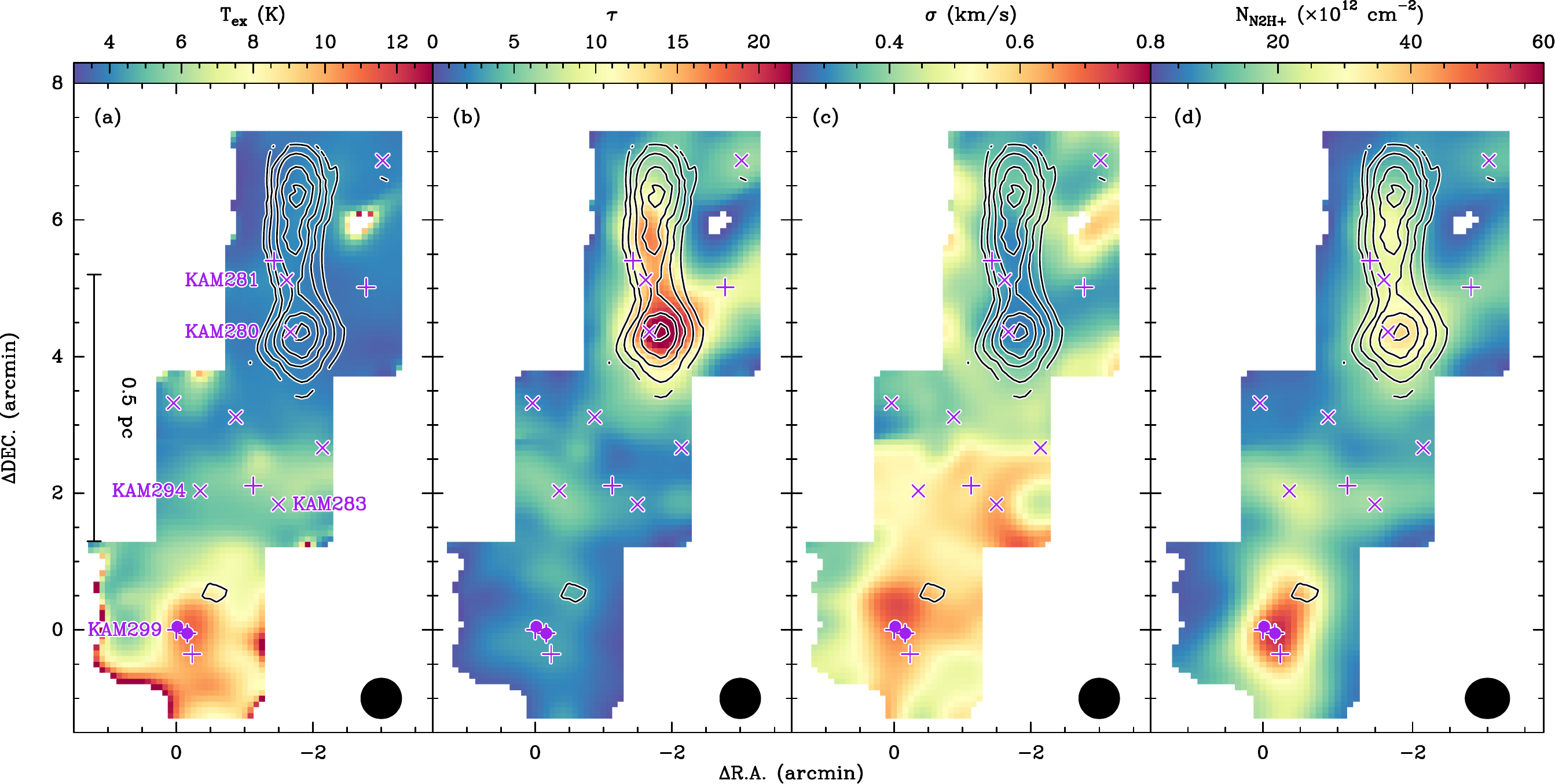}
\caption{{Distribution of (a) N$_{2}$H$^{+}$ (1--0) excitation temperature, (b) total optical depth, (c) velocity dispersion, and (d) N$_{2}$H$^{+}$ column density toward Serpens South. The red pixels at the southern boundary of panel a have large uncertainties in the fitted excitation temperatures. The overlaid HCNH$^{+}$ (2--1) integrated-intensity contours are the same as in Fig.~\ref{Fig:sers-hcnhp}a.}\label{Fig:n2hp-sers}}
\end{figure*}

Based on the HFS fit to the N$_{2}$H$^{+}$ (1--0) spectra, we find that the total optical depth of N$_{2}$H$^{+}$ (1--0) is generally larger than unity in our mapped regions. Therefore, we first apply the HFS fit to the N$_{2}$H$^{+}$ (1--0) data cube with the PySpecKit package \citep{2011ascl.soft09001G} to obtain the excitation temperature, total optical depth, velocity centroid, and velocity dispersion. We discard the fitted results for spectra with excitation temperatures lower than 5$\sigma$ in order to have reliable measurements of the derived column densities. The results are shown in Figs.~\ref{Fig:n2hp-serf} and \ref{Fig:n2hp-sers}. We find that the N$_{2}$H$^{+}$ (1--0) excitation temperatures are generally lower than 7~K in the Serpens filament, and 3.3--11.1~K in Serpens South. The highest excitation temperature of 11.1~K is found toward SSC. This is expected, because SSC tends to have the highest kinetic temperature and H$_{2}$ column density (see Fig.~\ref{Fig:sers-hcnhp}). The total optical depths of N$_{2}$H$^{+}$ (1--0) become quite high toward bolo4 in the Serpens filament and SSN in Serpens South, with the highest total optical depths reaching up to $\sim$14.5 and $\sim$24.5 in bolo4 and SSN, respectively. Figures~\ref{Fig:n2hp-serf}c and \ref{Fig:n2hp-sers}c present the distributions of the velocity dispersions of the N$_2$H$^+$ emission, which further support our hypothesis that the HCNH$^{+}$ emission is more prominent toward the quiescent regions that have velocity dispersions $\lesssim$0.3~\kms.

With the fitted values, the N$_{2}$H$^{+}$ column densities can be estimated using Eq.~(95) in \citet{2016PASP..128b9201M},
\begin{equation}\label{f.n2h+}
\begin{split}
 N_{\rm tot} =  \frac{3h}{8\pi^{3}\mu^{2}S} \frac{Q(T_{\rm ex})}{g_{J}} {\rm exp}(\frac{E_{\rm u}}{kT_{\rm ex}})[{\rm exp}(\frac{h\nu}{kT_{\rm ex}})-1]^{-1} 
             \times \int \tau{\rm d}\varv  \;.
\end{split}
\end{equation}
where $h$ is the Planck constant, $\mu$ is the dipole moment (see Table~\ref{Tab:lin}), $S$ is the line strength, $Q(T_{\rm ex})$ is the molecular partition function at the excitation temperature, $T_{\rm ex}$, $g_{J}$ describes the rotational degeneracy, $E_{\rm u}$ is the upper level energy of the transition, $T_{\rm bg}$ is the background temperature that is set to be 2.73~K here \citep{2009ApJ...707..916F}, and $\int \tau {\rm d}\varv$ is the integrated optical depth. For linear molecules, the line strength of a transition is given as $S = \frac{J_{\rm u}}{2J_{\rm u}+1}$ and the rotational degeneracy is given as $g_{J}= 2J_{\rm u}+1$, where $J_{\rm u}$ is the total angular momentum of the upper level of a given transition in a two-level system. The partition function of linear molecules can be approximated as $Q(T_{\rm ex})= \frac{kT_{\rm ex}}{hB}+\frac{1}{3}$, where $B$ is the rotational constant. 

With this equation, we are able to derive the column densities of N$_{2}$H$^{+}$. 
The results are shown in Figures~\ref{Fig:n2hp-serf}d and \ref{Fig:n2hp-sers}d. The derived N$_{2}$H$^{+}$ column densities are (0.4--10.8)$\times 10^{12}$~cm$^{-2}$ and (0.8--57.1)$\times 10^{12}$~cm$^{-2}$ for the Serpens filament and Serpens South, respectively. It is worth noting that the distribution of N$_{2}$H$^{+}$ column densities is slightly different from the corresponding N$_{2}$H$^{+}$ (1--0) integrated intensity map, which is caused by the optical depths of N$_{2}$H$^{+}$~(1--0) and the variations of excitation temperature. Furthermore, the HCNH$^{+}$ emission peaks coincide with the peaks of N$_{2}$H$^{+}$ column densities toward bolo2, bolo4, bolo12, and the peak in SSN2 in Figs.~\ref{Fig:n2hp-serf}d and \ref{Fig:n2hp-sers}d, which implies that HCNH$^{+}$ can, similarly to N$_{2}$H$^{+}$, trace prestellar core centers in such regions.

Based on the HFS fit to H$^{13}$CN (1--0), we find that its total optical depths can reach up to about 3. However, the HFS fitting toward the whole data cube tends to have large uncertainties for most pixels, which makes most of the fitting results unreliable. This could be caused by the anomalous line intensities of the hyperfine lines \citep[e.g.,][]{1982ApJ...258L..75W,1984A&A...139L..13C,2012MNRAS.420.1367L,2015ApJS..219....2J,2022A&A...658A..28G} and the relatively low signal-to-noise ratios. In order to balance the signal-to-noise ratio and optical depth effect, we use the $F=2-1$ satellite line (5/9 of the total optical depths) of H$^{13}$CN (1--0) to estimate H$^{13}$CN column densities in the optically thin approximation. 

As for HCNH$^{+}$, the peak intensities of its (2--1) and (3--2) lines are typically $<$0.5~K. Adopting a beam dilution factor of unity and an excitation temperature of 12~K, their peak optical depths are lower than 0.1 according to the radiative transfer equation. This suggests that HCNH$^{+}$ (2--1) and (3--2) are optically thin. 

\begin{figure}[!htbp]
\centering
\includegraphics[width = 0.45 \textwidth]{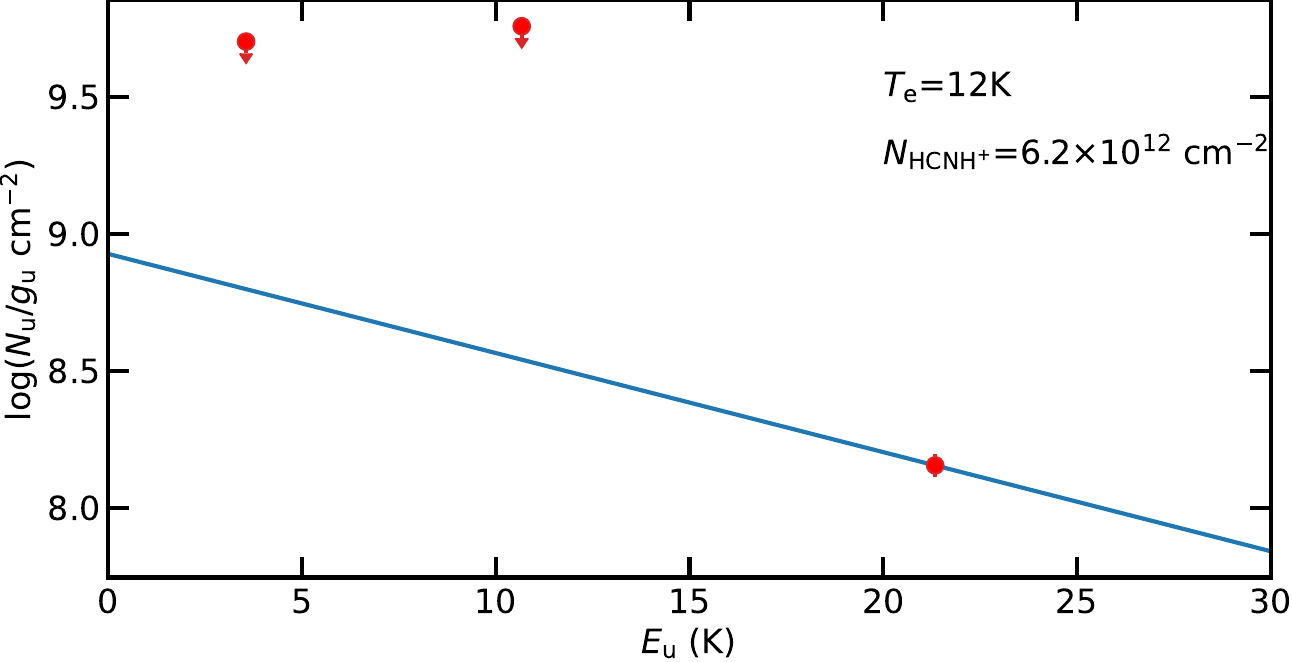}
\caption{{Population diagram of HCNH$^{+}$ toward KAM299. The blue line represents the fit expected from the excitation temperature and HCNH$^{+}$ column density which are indicated in the top right corner.  
}\label{Fig:rd}}
\end{figure}

We performed a similar estimate for HN$^{13}$C (1--0) and H$^{13}$CO$^{+}$(1--0). We find that these lines' peak optical depths can be higher than unity. HN$^{13}$C (1--0) also has HFS components, but these components are too close in frequency to allow for an accurate determination of its optical depth. Hence, HN$^{13}$C (1--0) and H$^{13}$CO$^{+}$(1--0) were simply assumed to be optically thin in this study. Assuming conditions of local thermodynamic equilibrium (LTE) and using the optically thin approximation, we can therefore estimate the beam-averaged column densities of these molecules. Nevertheless, we must exercise caution when interpreting the resulting H$^{13}$CN, HN$^{13}$C, and H$^{13}$CO$^{+}$ column densities in regions with high peak intensities, as they can be only taken as the lower limits in case of high optical depths. 
Excitation temperatures are needed to determine the molecular column densities. In the optically thin regime, the HCNH$^{+}$ excitation temperature can be estimated from the line ratio between HCNH$^{+}$ (2--1) and (3--2). Because of the relatively low signal-to-noise ratios in the HCNH$^{+}$ images, we only estimate the excitation temperatures toward emission peaks. Toward bolo12 and bolo4 in the Serpens filament, we obtain excitation temperatures of $\sim$7~K and $\sim$18~K, respectively. Toward KAM280, we convolve the HCNH$^{+}$ (2--1) raw data of Serpens South to an angular resolution of 27\arcsec\, to match the angular resolution of the HCNH$^{+}$ (3--2) spectra, and then derive the line ratio which indicates an excitation temperature of $\sim$12~K. The integrated intensity ratio between HCNH$^{+}$ (2--1) and HCNH$^{+}$ (1--0) is $\sim$2 in the averaged spectra (see Fig.~\ref{Fig:avsp}), corresponding to an excitation temperature of $\sim$10~K. The excitation temperatures are close to the gas kinetic temperature derived from ammonia \citep[e.g.,][]{2016ApJ...833..204F}. These estimates imply that the excitation temperatures might vary within the range of 7--18~K. In the following analysis, we assume a constant excitation temperature of 12~K in order to derive the column density of HCNH$^{+}$. Considering an excitation temperature range of 7--18~K, the assumption of a fixed excitation temperature can introduce uncertainties in the derived HCNH$^{+}$ column densities of at most 15\%. Making use of the excitation temperature, we made a fit to the HCNH$^{+}$ (3--2) data of KAM299 in Fig.~\ref{Fig:rd}, which demonstrates that our sensitivities are not sufficient to detect HCNH$^{+}$ (1--0) and (2--1). 

Because H$^{13}$CN (1--0), HN$^{13}$C (1--0), and H$^{13}$CO$^{+}$(1--0) have quite high critical densities (see Table~\ref{Tab:lin}), these transitions are likely sub-thermal. Hence, we adopt a lower excitation temperature of 5~K for H$^{13}$CN (1--0), HN$^{13}$C (1--0), and H$^{13}$CO$^{+}$(1--0). This assumption is consistent with the excitation temperature assumed for these species in previous studies toward dark clouds \citep[e.g.,][]{1998ApJ...503..717H}. If the excitation temperature is as high as 10 K, the derived column densities are underestimated by only $\sim$6\%\,for H$^{13}$CN, HN$^{13}$C, and H$^{13}$CO$^{+}$. 
Using the assumed excitation temperatures, we can calculate the column densities of HCNH$^{+}$, H$^{13}$CN, HN$^{13}$C, and H$^{13}$CO$^{+}$ from HCNH$^{+}$ (2--1), H$^{13}$CN (1--0), HN$^{13}$C (1--0), and H$^{13}$CO$^{+}$(1--0), respectively. The distributions of these molecular column densities are shown in Appendix~\ref{app.ncol}.
The derived column densities are listed in Table~\ref{Tab:column}, where the errors were estimated through a Monte Carlo analysis. In this analysis, we estimate the uncertainties in the column densities derived by generating 10,000 realizations randomly sampled from the Gaussian distributions associated with integrated intensities, optical depths, and line widths, propagating these samples through through Eq.~(\ref{f.n2h+}), and then analyzing the resulting distribution of outcomes to estimate the uncertainties.





\begin{figure*}[!htbp]
\centering
\includegraphics[width = 1 \textwidth]{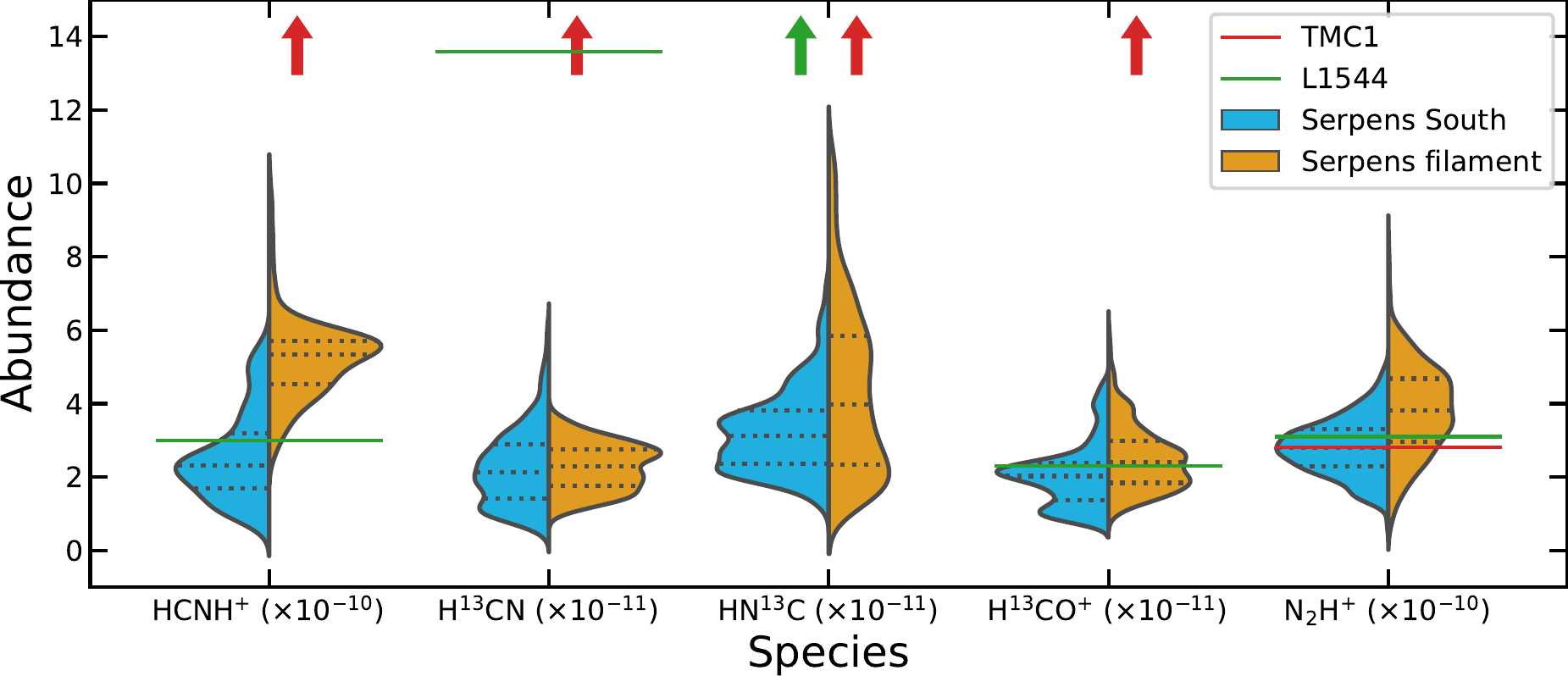}
\caption{{Comparison of the fractional abundances of different species with respect to H$_{2}$ determined on a pixel-by-pixel basis. The y-axis has been divided by the scaling values which are given in parentheses after the molecule names along the x-axis for better visualization. The dotted lines indicate the 25\%, 50\%, and 75\% percentiles of the respective distributions. The molecular abundances in TMC1 and L1544 are also indicated for comparison. The arrows indicate that their abundances are higher than the maximum value in the y-axis.  
}\label{Fig:abundance}}
\end{figure*}

\begin{figure*}[!htbp]
\centering
\includegraphics[width = 1 \textwidth]{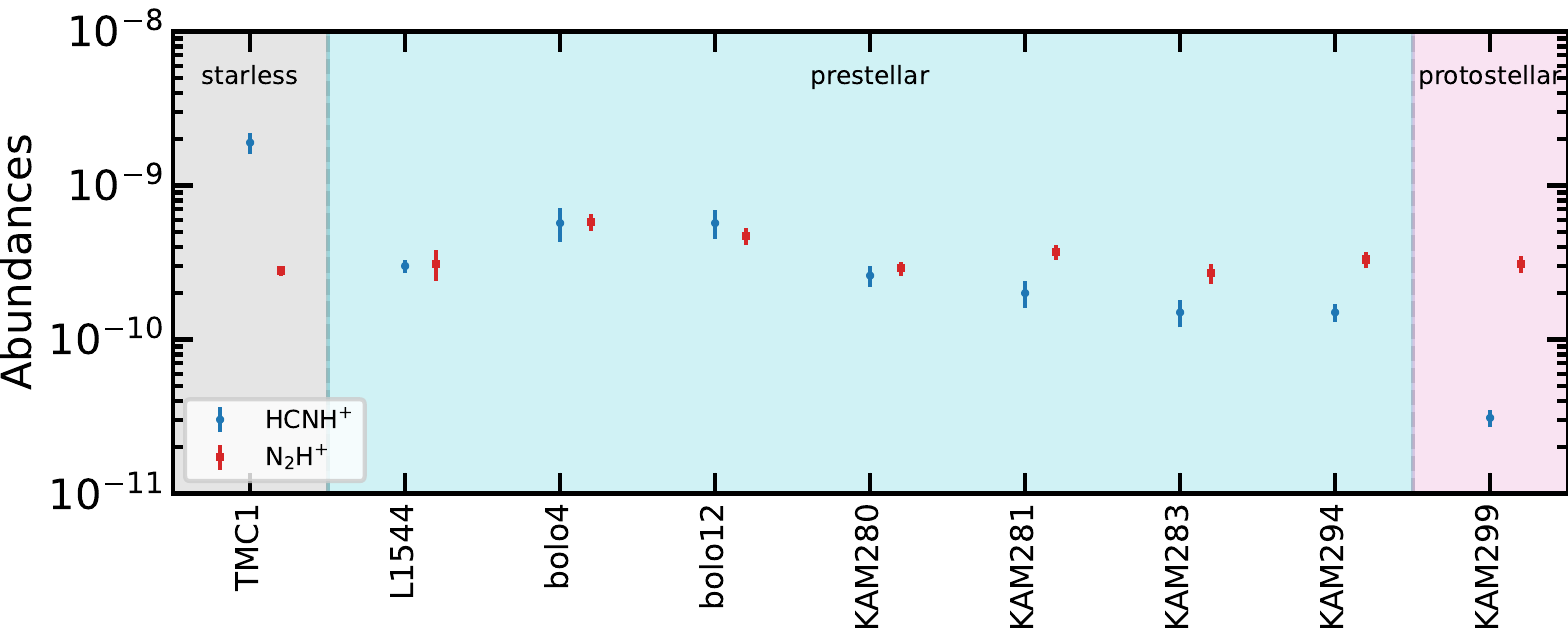}
\caption{{HCNH$^{+}$ (blue) and N$_{2}$H$^{+}$ (red) abundances relative to H$_{2}$ as a function of evolutionary stages. 
}\label{Fig:stage}}
\end{figure*}

\begin{figure*}[!htbp]
\centering
\includegraphics[width = 1 \textwidth]{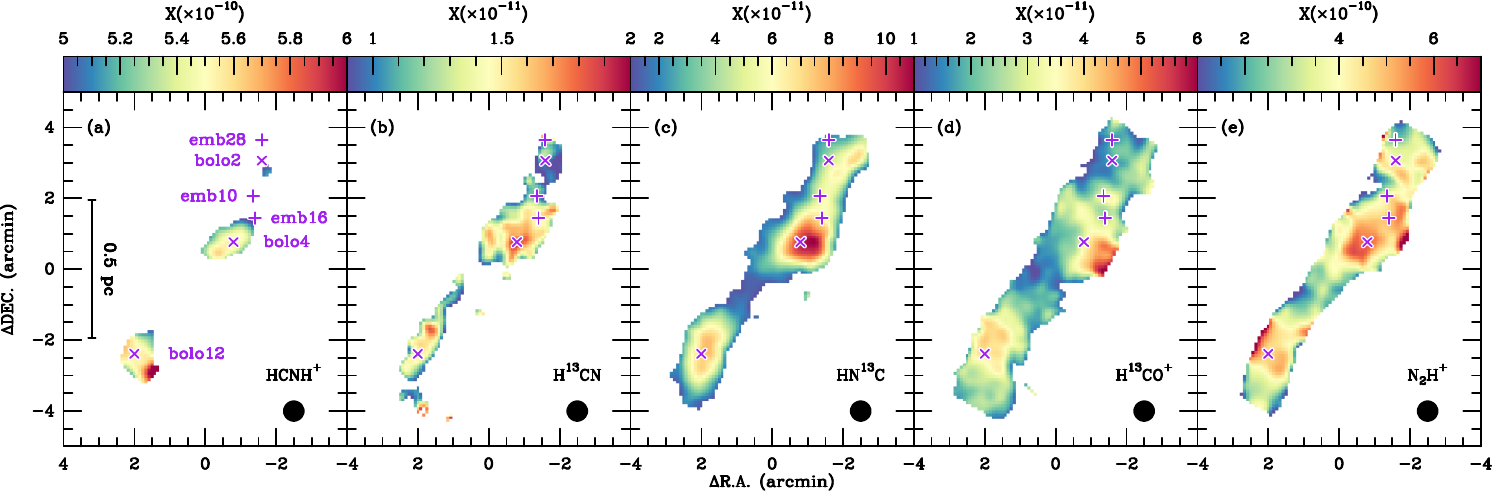}
\caption{{Molecular abundance distributions of HCNH$^{+}$ (panel~a), H$^{13}$CN (panel~b), HN$^{13}$C (panel~c), H$^{13}$CO$^{+}$ (panel~d), and N$_{2}$H$^{+}$ (panel~e) in the Serpens filament. The beam size is shown in the lower right corner of each panel. In all panels, the (0, 0) offset corresponds to $\alpha_{\rm J2000}$=18$^{\rm h}$28$^{\rm m}$50$\rlap{.}^{\rm s}$4, $\delta_{\rm J2000}$=00$^{\circ}$49$^{\prime}$58$\rlap{.}^{\prime \prime}$72. The markers are the same as in Fig.~\ref{Fig:sf-hcnh+}.}\label{Fig:sf-abun}}
\end{figure*}

\begin{figure*}[!htbp]
\centering
\includegraphics[width = 1 \textwidth]{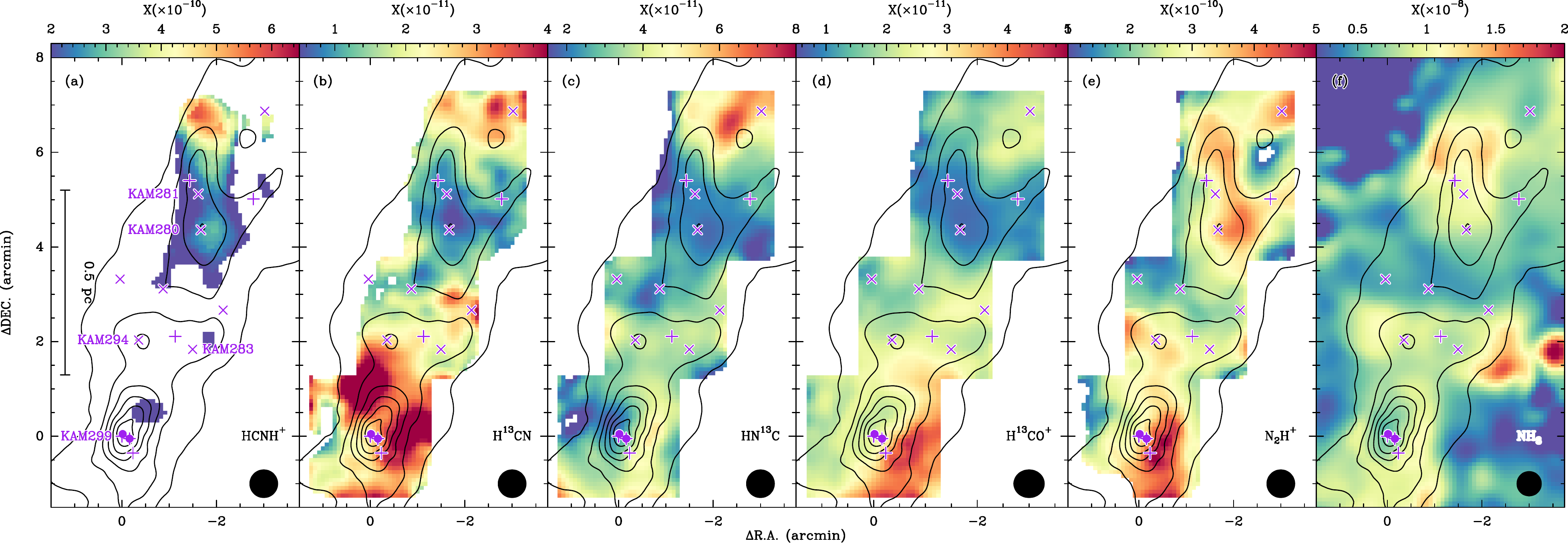}
\caption{{Molecular abundance distributions of HCNH$^{+}$ (panel a), H$^{13}$CN (panel b), HN$^{13}$C (panel c), H$^{13}$CO$^{+}$ (panel d), N$_{2}$H$^{+}$ (panel e), and para-NH$_{3}$ (panel f, from \citealt{2016ApJ...833..204F}) in Serpens South. The overlaid \textit{Herschel} dust-based H$_{2}$ column density contours start at 2.4$\times 10^{22}$~cm$^{-2}$ and increase by 2.4$\times 10^{22}$~cm$^{-2}$. The beam size is shown in the lower right corner of each panel. In all panels, the (0, 0) offset corresponds to $\alpha_{\rm J2000}$=18$^{\rm h}$30$^{\rm m}$04$\rlap{.}^{\rm s}$19, $\delta_{\rm J2000}$=$-$02$^{\circ}$03$^{\prime}$05$\rlap{.}^{\prime \prime}$5. The markers are the same as in Fig.~\ref{Fig:sers-hcnhp}.}\label{Fig:sers-abun}}
\end{figure*}

Molecular abundances, $X({\rm A})=N({\rm A})/N({\rm H}_{2})$, are derived from the ratio between molecular column densities, $N({\rm A})$, and H$_{2}$ column densities, $N({\rm H}_{2})$, by adopting the \textit{Herschel} dust-based H$_{2}$ column density maps \citep{2015A&A...584A..91K,2021MNRAS.500.4257F}. The derived molecular abundances are listed in Table~\ref{Tab:abun}, and the statistical results of the derived molecular abundances are presented in Fig.~\ref{Fig:abundance}. 
Compared to TMC1 and L1544 \citep{2002ApJ...572..238C,2002ApJ...565..344C,2013ChRv..113.8710A,2017MNRAS.470.3194Q,1998ApJ...503..717H}, we find that the H$^{13}$CN, HN$^{13}$C, and H$^{13}$CO$^{+}$ abundances are lower in Serpens South and the Serpens filament (see Fig.~\ref{Fig:abundance}). On the other hand, the derived N$_{2}$H$^{+}$ abundances in TMC1 and L1544 are comparable to those of our mapped regions, which suggests that N$_{2}$H$^{+}$ abundances do not vary much in different environments. For HCNH$^{+}$, the molecular abundances vary from $6.5\times 10^{-11}$ to $5.9\times 10^{-10}$ with a median value of 2.4$\times 10^{-10}$ and 5.6$\times 10^{-10}$ in Serpens South and the Serpens filament, respectively. These values are comparable to that in L1544 \citep[$\sim$3$\times 10^{-10}$,][]{2017MNRAS.470.3194Q} but lower than that of TMC1 \citep[$\sim$1.9$\times 10^{-9}$,][]{1991A&A...247..487S} and generally higher than the abundances measured in high-mass star formation regions \citep[0.9--14$\times 10^{-11}$,][]{2021A&A...651A..94F}. 

We further investigate the HCNH$^{+}$ abundance variation as a function of evolutionary stage in Fig.~\ref{Fig:stage}, which suggests that the HCNH$^{+}$ abundance decreases by almost two orders of magnitude from the starless phase to the protostellar phase. This finding is consistent with the results toward high-mass star formation regions where the decrease in the HCNH$^{+}$ abundance is attributed to an evolutionary effect \citep{2021A&A...651A..94F}.





The spatial distributions of molecular fractional abundances are shown in Figs.~\ref{Fig:sf-abun} and \ref{Fig:sers-abun}.
In Fig.~\ref{Fig:sf-abun}, H$^{13}$CO$^{+}$ abundances peak toward the west of the dust continuum peak in bolo4.
In contrast, HCNH$^{+}$, HN$^{13}$C, H$^{13}$CN, and N$_{2}$H$^{+}$ can still trace the core center of bolo4. It is expected that H$^{13}$CO$^{+}$ would not be a reliable tracer of the core center in bolo4, as CO, which is a main precursor of HCO$^{+}$, has been observed to undergo depletion in this region \citep{2018A&A...620A..62G,2021A&A...646A.170G}. This is consistent with the fact that CO freezes out onto dust grains for cold and dense regions.

In Fig.~\ref{Fig:sers-abun}, we find that the molecular fractional abundances of HCNH$^{+}$, H$^{13}$CN, HN$^{13}$C, H$^{13}$CO$^{+}$ are lower in the southern part of SSN2 than the northern part of SSN2 by a factor of $\gtrsim$2, which suggests a north-south abundance gradient across SSN2. The H$^{13}$CN, HN$^{13}$C, H$^{13}$CO$^{+}$ abundances are even lower in SSN1. This could be caused by the freeze-out process on to dust grains that affects these molecules or their main precursor molecules. For H$^{13}$CN, HN$^{13}$C, H$^{13}$CO$^{+}$, the depletion sizes are found to be $\sim$0.3~pc. In contrast, N$_{2}$H$^{+}$ and NH$_{3}$ appear to still be abundant in SSN, which supports the selective freeze-out whereby molecules exhibit different behaviors when interacting with dust grains \citep[e.g.,][]{2007ARA&A..45..339B}. Our results are roughly in line with the scenario that CO is the first to be depleted, followed by HCN, HNC, and HCNH$^{+}$, while NH$_{3}$ and N$_{2}$H$^{+}$ are least affected. We also notice that even N$_{2}$H$^{+}$ abundances appear to drop toward the peak of the SSC when compared to ambient gas, which is similar to NH$_{3}$ \citep{2016ApJ...833..204F}. Such low abundances indicate that even N$_{2}$H$^{+}$ and NH$_{3}$ begin to deplete from the gas phase in SSC  \citep{2002ApJ...570L.101B,2004A&A...419L..35B,2007ARA&A..45..339B}. Another potential scenario is that the elevated kinetic temperatures within the SSC lead to CO desorption back to the gas phase, facilitating the efficient destruction of N$_{2}$H$^{+}$. This phenomenon would also trigger the desorption of NH${3}$ back to the gas phase, potentially enriching its abundances. However, the NH$_{3}$ abundance remains relatively low within the SSC, which might be attributed to the inefficient NH$_{3}$ desorption.

\begin{figure*}[!htbp]
\centering
\includegraphics[width = 1 \textwidth]{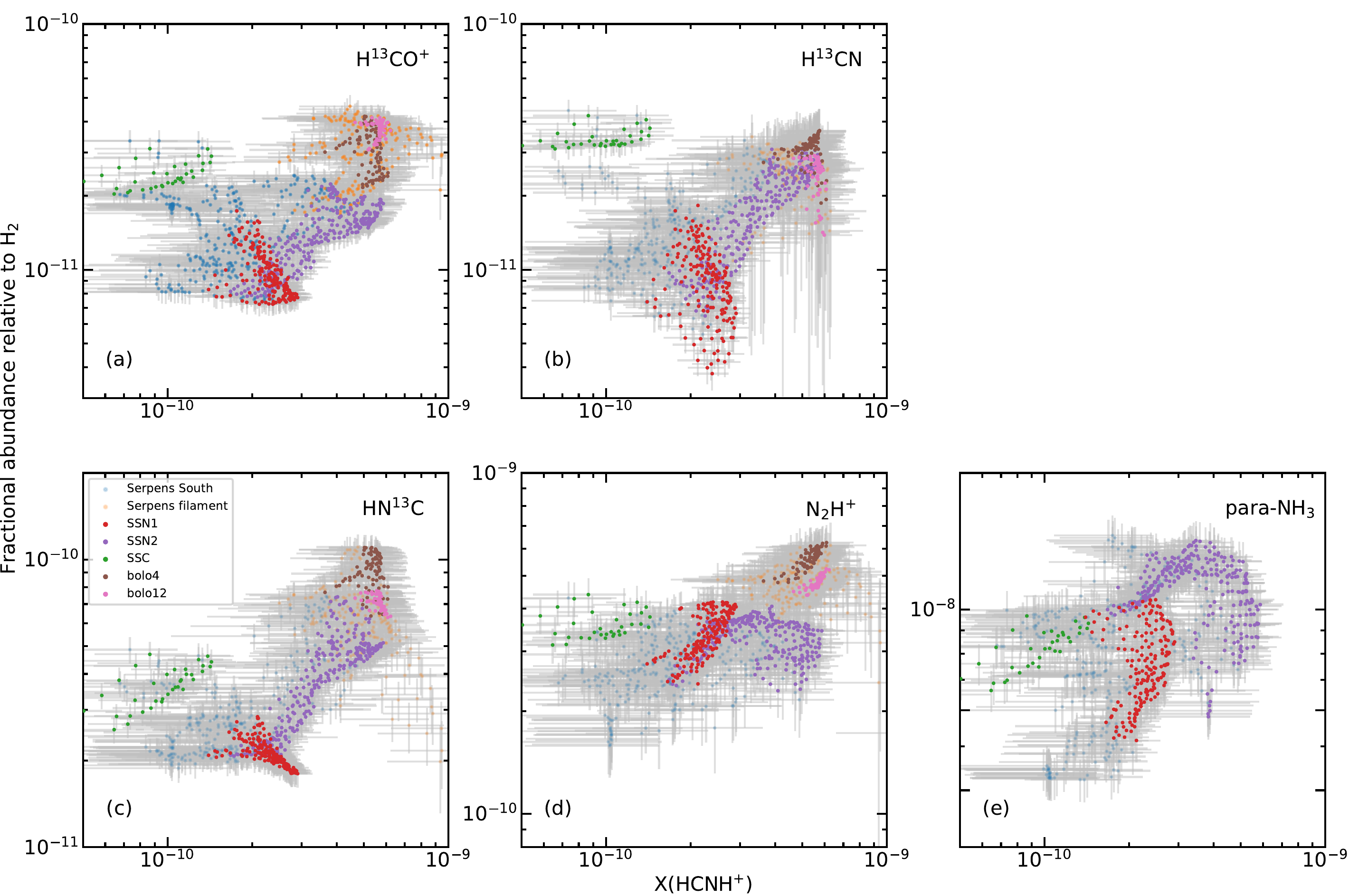}
\caption{{Pixel-by-pixel comparison of the abundances of HCNH$^{+}$ and the other five molecules in the Serpens filament and Serpens South. The corresponding molecule is indicated in the upper right corner of each panel. The subregions are indicated by the different colors in the legend of panel~c.}\label{Fig:ppabun}}
\end{figure*}

Figure~\ref{Fig:ppabun} shows the pixel-by-pixel comparison of the abundances of HCNH$^{+}$ and the other five molecules. It is evident that H$^{13}$CO$^{+}$, H$^{13}$CN, and HN$^{13}$C abundances generally increase with increasing HCNH$^{+}$ abundances toward SSN2, SSC, bolo4, and bolo12. This can be readily explained by the freeze-out of CO, HCN, and HNC. However, we find that SSN1 exhibits the opposite trend where the H$^{13}$CO$^{+}$, H$^{13}$CN, and HN$^{13}$C abundances generally decrease with increasing HCNH$^{+}$ abundances. Instead, HCNH$^{+}$ abundances are positively correlated with N$_{2}$H$^{+}$ and NH$_{3}$ in SSN1, in agreement with the spatially coincident distribution of HCNH$^{+}$ and N$_{2}$H$^{+}$ as shown in Fig.~\ref{Fig:sers-abun}. This implies  a different chemical formation pathway for HCNH$^{+}$ in SSN1 (details are discussed in Sect.~\ref{sec.chemistry}). We also find that SSC forms a distinct group in this figure, especially from the comparison of HCNH$^{+}$ with H$^{13}$CO$^{+}$ and H$^{13}$CN. The abundance ratios of $X({\rm HCNH}^{+})/X({\rm H^{13}CO}^{+})$ and $X({\rm HCNH}^{+})/X({\rm H^{13}CN})$ are $\lesssim$5 in SSC, which are much lower than in other regions. This trend is similar to the results of \citet{2021A&A...651A..94F} where warmer sources have lower $X({\rm HCNH}^{+})/X({\rm HCO}^{+})$ and $X({\rm HCNH}^{+})/X({\rm HCN})$ ratios. This can be attributed to the elevated kinetic temperatures which in turn enhance the H$^{13}$CO$^{+}$ and H$^{13}$CN abundances.

\begin{table*}[!hbt]
\caption{Molecular column densities of the observed targets.}\label{Tab:column}
\small
\centering
\begin{tabular}{ccccccccc}
\hline \hline
Name             & $T_{\rm d}$ & $T_{\rm g}$ & $N$(H$_{2}$) & $N$(HCNH$^{+}$)$^{*}$ & $N$(H$^{13}$CN)$^{*}$  &  $N$(HN$^{13}$C)$^{*}$ &  $N$(H$^{13}$CO$^{+}$)$^{*}$   & $N$(N$_{2}$H$^{+}$) \\ 
                  & (K)  & (K) & ($\times 10^{22}$~cm$^{-2}$) &   ($\times 10^{12}$~cm$^{-2}$)  &  ($\times 10^{11}$~cm$^{-2}$)  &    ($\times 10^{11}$~cm$^{-2}$)           &    ($\times 10^{11}$~cm$^{-2}$) &  ($\times 10^{12}$~cm$^{-2}$)  \\ 
(1)              & (2)               & (3)              & (4)              & (5)    & (6)      & (7)  & (8)  & (9)    \\
\hline
The Serpens filament &  12.0-16.3               & ...             & 0.3-2.2 & 5.1--10.3               & 1.1--6.3               & 0.7--18.2              & 0.6--7.0    & 0.4--10.8     \\
\hline
bolo4      &  12.4  & ... & 1.7  & 9.7$\pm$2.1 &  6.1$\pm$1.1 & 17.9$\pm$0.3  &  5.8$\pm$0.2   &  10.0$\pm$0.7 \\
bolo12     &  12.5  & ... & 1.8  & 10.1$\pm$1.8 & 4.7$\pm$0.5 & 12.9$\pm$0.1  &  6.7$\pm$0.2   &  8.3$\pm$0.5  \\
\hline
Serpens South     &  11.1-16.5               & 10.3--17.5            & 1.8--16.7    & 4.2--27.1               & 2.5--46.8              & 1.9--45.0              & 1.1--41.2           & 0.8--57.1     \\
\hline
KAM280     & 11.4 & 10.9 & 9.6 & 24.8$\pm$2.3 & 6.3$\pm$0.3 & 18.7$\pm$0.3  & 7.5$\pm$0.2  & 35.1$\pm$2.0   \\
KAM281     & 11.6 & 10.7 & 8.6 & 17.1$\pm$2.4 & 6.7$\pm$0.4 & 18.0$\pm$0.3  & 6.8$\pm$0.1  & 24.7$\pm$1.5       \\
KAM283    & 12.4 & 12.6 & 6.6 & 10.8$\pm$1.8 & 12.9$\pm$0.4  & 23.2$\pm$0.5    & 14.6$\pm$0.3 & 17.7$\pm$1.4     \\
KAM294    & 13.4 & 12.9 & 7.3 & 12.6$\pm$1.3 & 20.1$\pm$0.4 & 26.8$\pm$0.5    & 16.2$\pm$0.2 & 23.9$\pm$1.4     \\
KAM299    & 16.0 & 16.4 & 16.4 & 6.2$\pm$0.5 & 42.2$\pm$0.3 & 37.3$\pm$0.9    & 37.3$\pm$0.9 & 50.6$\pm$2.2  \\
\hline
\end{tabular}
\tablefoot{(1) Source name. (2) Dust temperature. (3) Gas kinetic temperature. ``..." indicates that no information is available. (4) H$_{2}$ column density. (5) HCNH$^{+}$ column density. For KAM283, KAM294, and KAM299 where HCNH$^{+}$~(2--1) emission is not detected, we use HCNH$^{+}$~(3--2) to estimate the column densities at a resolution of 27\arcsec. (6) H$^{13}$CN column density. (7) HN$^{13}$C column density. (8) H$^{13}$CO$^{+}$ column density. (9) N$_{2}$H$^{+}$ column density. ``$^{*}$" indicates that these column densities are derived under the optically thin assumption. A constant excitation temperature of 12~K is assumed for HCNH$^{+}$, while a constant excitation temperature of 5~K is assumed for H$^{13}$CN, HN$^{13}$C, and H$^{13}$CO$^{+}$.}
\normalsize
\end{table*}

\begin{table*}[!hbt]
\caption{Molecular abundances with respect to H$_{2}$ toward different targets.}\label{Tab:abun}
\normalsize
\centering
\begin{tabular}{cccccc}
\hline \hline
Name             & $X$(HCNH$^{+}$) & $X$(H$^{13}$CN)  &  $X$(HN$^{13}$C) &  $X$(H$^{13}$CO$^{+}$)   & $X$(N$_{2}$H$^{+}$) \\ 
                 &   ($\times 10^{-10}$)  &  ($\times 10^{-11}$)  &    ($\times 10^{-11}$)           &    ($\times 10^{-11}$) &  ($\times 10^{-10}$)  \\ 
(1)              & (2)               & (3)              & (4)              & (5)    & (6)        \\
\hline
The Serpens filament &  2.5--10.1 & 1.3--3.8 &  0.9--11.0 & (0.9--6.1)   & 0.6--8.6   \\
\hline
bolo4      & 5.7$\pm$1.4 & 3.6$\pm$0.8  & 10.5$\pm$0.1 & 3.4$\pm$0.4 &  $5.8\pm0.7$   \\
bolo12     & 5.7$\pm$1.2 & 2.7$\pm$0.4 &  7.2$\pm$0.8 & 3.8$\pm$0.4 &  $4.7\pm0.6$   \\
\hline
Serpens South & 0.3--5.9 & 0.4--3.7  & 0.9--7.4  & 0.7--4.7  &  0.5--5.2 \\
\hline
KAM280    &  2.6$\pm$0.4 & 0.7$\pm$0.1   & 2.0$\pm$0.2 & 0.8$\pm$0.1 & 2.9$\pm$0.3  \\
KAM281    &  2.0$\pm$0.4 & 0.8$\pm$0.1   & 2.1$\pm$0.2 & 0.8$\pm$0.1 & 3.7$\pm$0.4  \\
KAM283    &  1.5$\pm$0.3 & 2.0$\pm$0.2 & 3.5$\pm$0.4 & 2.2$\pm$0.2 & 2.7$\pm$0.4   \\
KAM294    &  1.5$\pm$0.2 & 2.7$\pm$0.3 & 3.7$\pm$0.4 & 2.2$\pm$0.2 & 3.3$\pm$0.4 \\
KAM299    &  0.31$\pm$0.04  & 2.6$\pm$0.3 & 2.3$\pm$0.2 & 2.3$\pm$0.2 & 3.1$\pm$0.4 \\
\hline
TMC1      & 19.0$\pm$3.0    & 17.8$\pm$2.1 & 41.7$\pm$6.2 &  14.4$\pm$0.3 &  2.8$\pm$0.2  \\
L1544     & 3.0$\pm$0.3  & 13.6$\pm$1.7 &  31.4$\pm$5.7  & 2.3$\pm$0.7 &  3.1$\pm$0.7  \\
\hline
\end{tabular}
\tablefoot{(1) Source name. (2) HCNH$^{+}$ fractional abundance relative to H$_{2}$. (3) H$^{13}$CN fractional abundance relative to H$_{2}$. (4) HN$^{13}$C fractional abundance relative to H$_{2}$. (5) H$^{13}$CO$^{+}$ fractional abundance relative to H$_{2}$. (6) N$_{2}$H$^{+}$ fractional abundance relative to H$_{2}$. The uncertainties of molecular abundances are estimated by assuming 10\% uncertainties in the dust-based H$_{2}$ column densities \citep{2015A&A...584A..91K}.  The molecular abundances in TMC1 are taken from \citet{1997ApJ...486..862P} and \citet{1991A&A...247..487S}, while the molecular abundances in L1544 are based on \citet{2002ApJ...572..238C}, \citet{2002ApJ...565..344C}, \citet{2017MNRAS.470.3194Q} , and \citet{1998ApJ...503..717H}. }
\normalsize
\end{table*}


\section{Discussion}\label{Sec:dis}



\subsection{Environmental dependence}\label{sec.dependence}
As shown in Sect.~\ref{sec.col}, molecular abundances are sensitive to environmental conditions. Here, we attempt to quantify how the observed abundances depend on the physical parameters. 

\begin{figure*}[!htbp]
\centering
\includegraphics[width = 1 \textwidth]{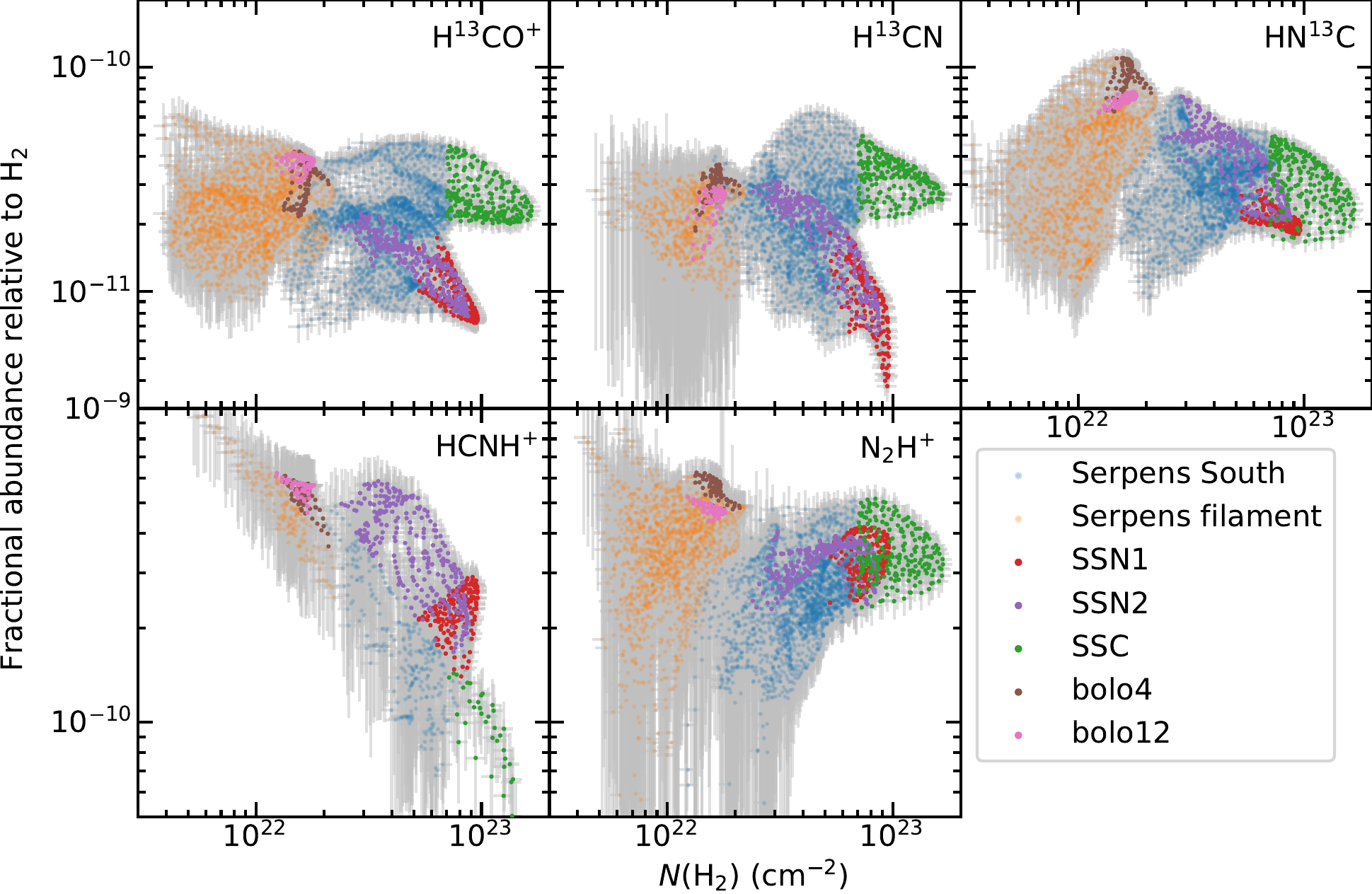}
\caption{{Fractional abundances as a function of H$_{2}$ column density for the five molecules indicated in the top right corner of each panel. Different regions are indicated by different colors as shown in the legend. 
}\label{Fig:abun-h2}}
\end{figure*}

Figure~\ref{Fig:abun-h2} presents the comparison between the derived molecular abundances and H$_{2}$ column densities for the five molecules. In contrast to other molecules, the abundance of HCNH$^{+}$ shows an overall anti-correlation with H$_{2}$ column density with a strongly negative Pearson correlation coefficient of $-$0.73. This further supports that HCNH$^{+}$ is abundant in low-density regions. It is also evident that the H$^{13}$CO$^{+}$, H$^{13}$CN, HN$^{13}$C, and HCNH$^{+}$ abundances decrease with increasing H$_{2}$ column density toward SSN. This pattern aligns with the depletion caused by the freeze-out process of CO, HCN, and HNC. Given that these molecules are known to be the main precursors of HCO$^{+}$ and HCNH$^{+}$, the freeze-out process causes the depletion of HCO$^{+}$ and HCNH$^{+}$. Interestingly, HCNH$^{+}$ has a lower abundance in SSC than in SSN. In contrast, HCO$^{+}$ and HCN are more abundant in SSC than in SSN, which could be potentially explained by the thermal desorption of CO, H$_{2}$O, and HCN from dust grains to the gas phase. Such desorption can be caused by elevated kinetic temperatures, for example due to outflow shocks. However, such a desorption process seems not to enhance the HCNH$^{+}$ abundance in SSC.

\begin{figure*}[!htbp]
\centering
\includegraphics[width = 1 \textwidth]{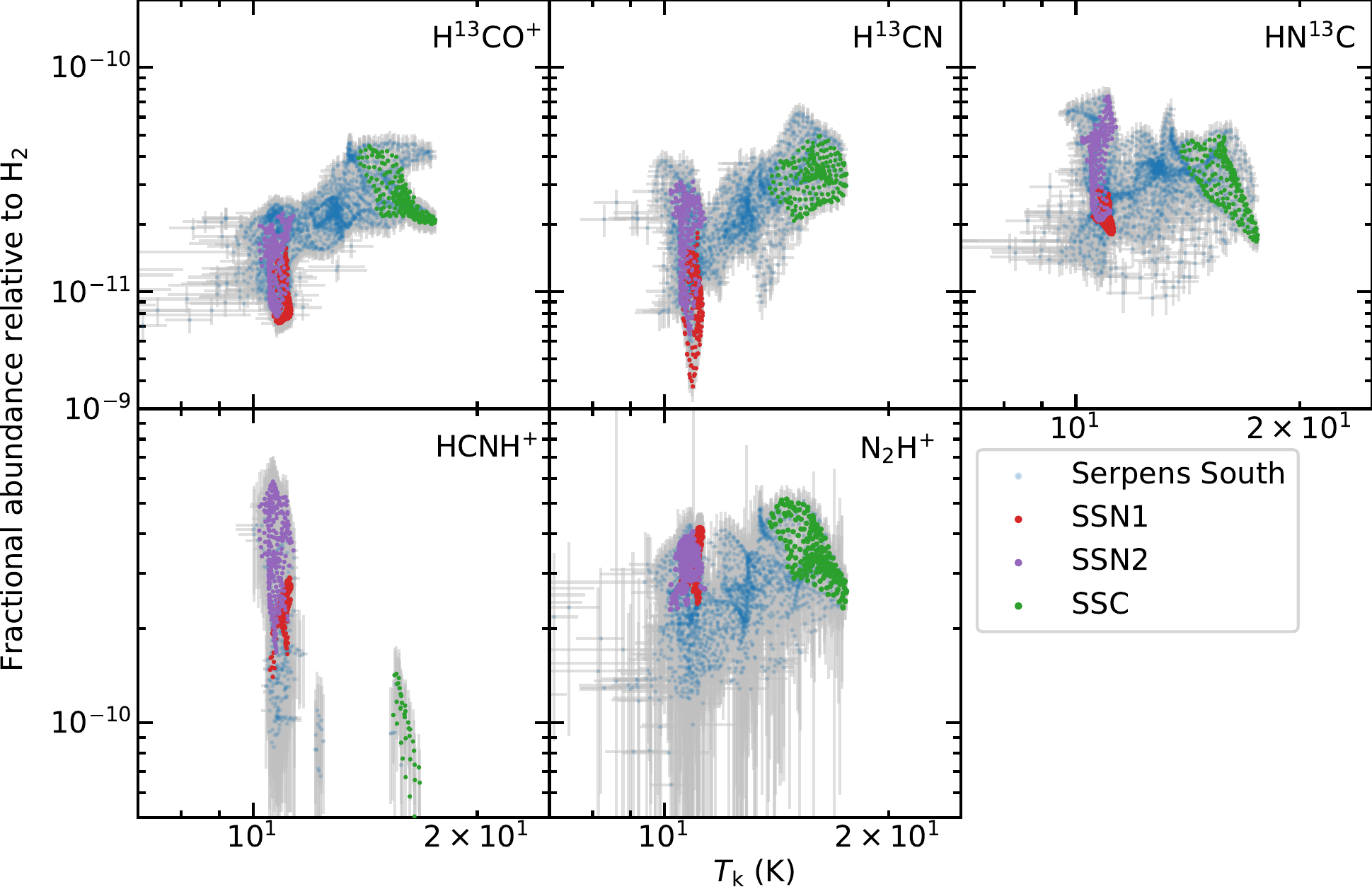}
\caption{{Same as Fig.~\ref{Fig:abun-h2} but as a function of the gas kinetic temperature derived from ammonia inversion transitions. 
}\label{Fig:abun-tk}}
\end{figure*}

In order to investigate the dependence of molecular abundances on gas kinetic temperature, we make use of the gas kinetic temperature derived from ammonia inversion transitions. Because of the lack of a gas kinetic temperature map toward the Serpens filament, we only investigate the molecular abundance variation as a function of the gas kinetic temperature toward Serpens South, which is shown in Fig.~\ref{Fig:abun-tk}. Overall, the H$^{13}$CO$^{+}$ and H$^{13}$CN abundances increase with increasing kinetic temperatures, while that of HCNH$^{+}$ shows an opposite trend. HN$^{13}$C and N$_{2}$H$^{+}$ appear to be the least affected by variation in the kinetic temperature. Toward the SSC, H$^{13}$CO$^{+}$, HN$^{13}$C, HCNH$^{+}$, and N$_{2}$H$^{+}$ exhibit a different behavior from H$^{13}$CN whose molecular abundances appear to increase with increasing kinetic temperature. The different behaviors between H$^{13}$CN and HN$^{13}$C could be explained by the fact that HCN/HNC abundance ratio increases with kinetic temperature \citep[e.g.,][]{2020A&A...635A...4H}. However, the gas kinetic temperature range (14--17 K) is too narrow to reach a more concrete conclusion.

\subsection{Comparison between chemical models and observations}\label{sec.chemistry}
As shown in Sect.~\ref{sec.dependence}, the HCNH$^{+}$ abundance appears to be anticorrelated with the H$_{2}$ column density and kinetic temperature, which indicates that the HCNH$^{+}$ abundance deficit could be caused by the increased H$_{2}$ number density and kinetic temperature. 

In order to test this hypothesis, we used the time dependent gas-grain chemical code, \textit{Chempl\footnote{\url{https://github.com/fjdu/chempl}}} \citep{2021RAA....21...77D}, to carry out astrochemical model calculations. The UMIST RATE12 chemical network\footnote{\url{http://udfa.ajmarkwick.net/}} is adopted for this study \citep{2013A&A...550A..36M}. Since HCN and HNC are important precursors of HCNH$^{+}$, their chemistry may also affect the chemistry of HCNH$^{+}$. 
Two reactions are believed to be important in the chemistry of HCN and HNC in molecular clouds \citep[e.g.,][]{1992A&A...256..595S,1996A&A...314..688T,2014ApJ...787...74G,2020A&A...635A...4H}:
\begin{equation}\label{f.h}
    \begin{split}
        {\rm HNC} + {\rm H} \to {\rm HCN} + {\rm H}
    \end{split}
\end{equation}
\begin{equation}\label{f.o}
    \begin{split}
        {\rm HNC} + {\rm O} \to {\rm NH} + {\rm CO}\;.
    \end{split}
\end{equation}
Previous studies suggest that the assumption of an energy barrier, $\Delta E_{10}$, of 200~K is suitable for reaction~(\ref{f.h}) \citep{2014ApJ...787...74G,2020A&A...635A...4H}, but quite different energy barriers $\Delta E_{11} =$ 20~K \citep{2020A&A...635A...4H} and $\Delta E_{11}=$ 1125~K \citep{2014ApJ...787...74G} have been proposed for reaction~(\ref{f.o}). Reaction~(\ref{f.h}) is included in the UMIST RATE12 chemical network, but the energy barrier is not up-to-date. On the other hand, reaction~(\ref{f.o}) is not included in the UMIST RATE12 chemical network. Hence, the two reactions with updated rate coefficients are augmented in the chemical network. The two different energy barriers of $\Delta E_{11}$ are used for comparison in the following. 


Initial conditions are needed in the chemical code for the calculations. The initial elemental abundances are the same as in Table~3 of \citet{2013A&A...550A..36M}, which are based on diffuse cloud values. The interstellar radiation field, $G_{0}$, is set to be $G_{0}=1$ in Habing units \citep{1978ApJS...36..595D}. Although previous studies suggest that the abundances of HCN, HNC, and HCNH$^{+}$ can be regulated by the cosmic-ray ionization rate \citep[CRIR;][]{2021A&A...651A..94F,2022ApJ...939..119B}, our measurements probe almost the same physical conditions and the CRIR is the same for the all of the cloud volumes studies by us. 
Hence, we 
adopt the canonical value of $1.36\times 10^{-17}$~s$^{-1}$ in dense molecular gas for the CRIR \citep[e.g.,][]{2000A&A...358L..79V}. As shown in Figs.~\ref{Fig:sf-hcnh+}--\ref{Fig:sers-hcnhp}, the HCNH$^{+}$ emitting regions have kinetic temperatures ranging from 10~K to 20~K and H$_{2}$ column densities of $>1\times 10^{22}$~cm$^{-2}$ (i.e., $A_{\rm v}\gtrsim$10~mag). We use two values from the kinetic temperature, 10~K and 20~K, to investigate the effects of the kinetic temperature variations. Given their high H$_{2}$ column densities, our target regions are well shielded from Ultraviolet radiation.  We use $A_{\rm v}=10$~mag as a fiducial case. The modeled H$_{2}$ number densities are in the range of 10$^{2}$--10$^{6}$~cm$^{-3}$. 

\begin{figure*}[!htbp]
\centering
\includegraphics[width = 0.95 \textwidth]{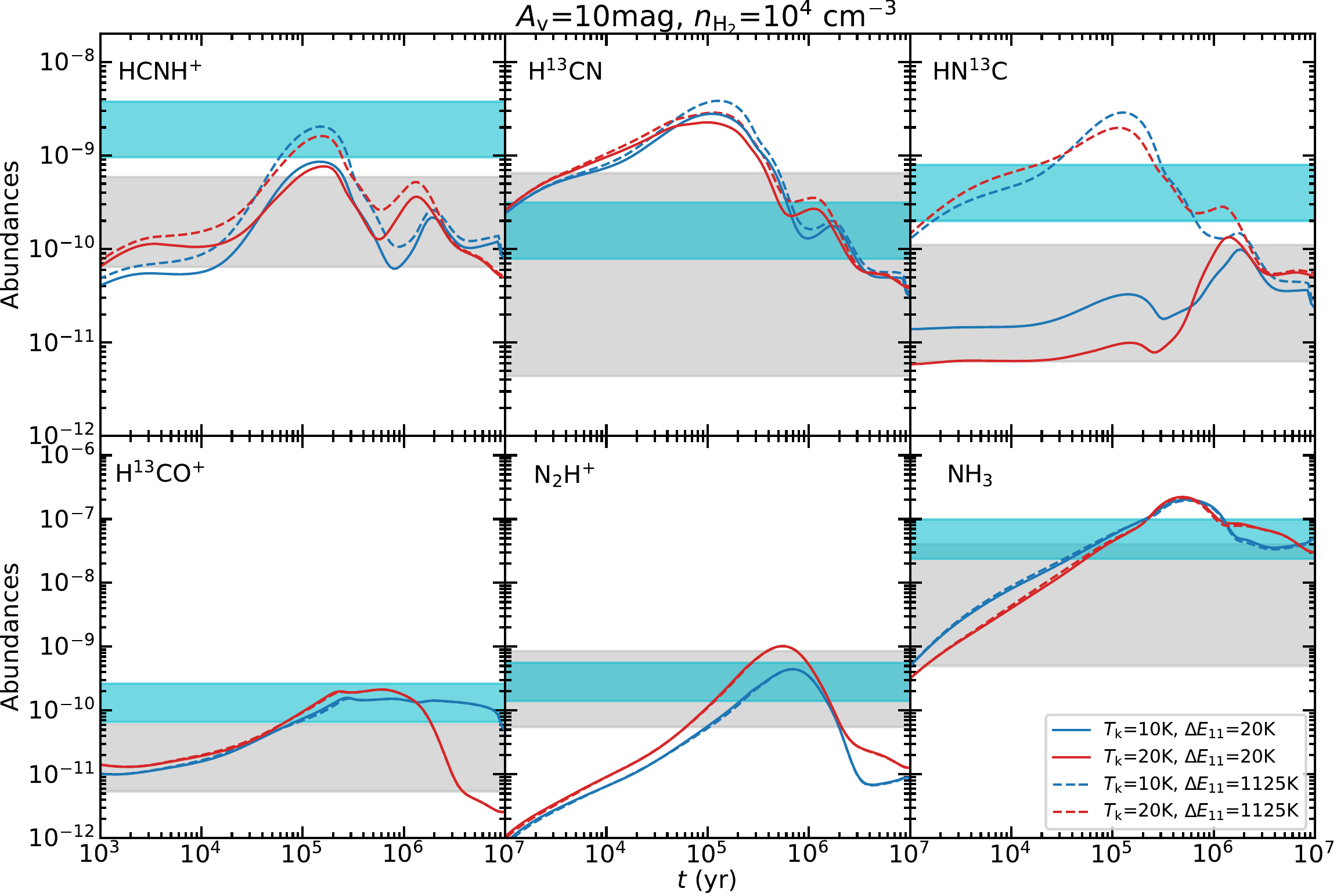}
\caption{{Molecular abundances relative to H$_{2}$ as a function of time, estimated from \textit{Chempl} \citep{2021RAA....21...77D}. The different colors correspond to different kinetic temperatures. The solid and dashed lines represent the modeling results with the energy barrier $\Delta E_{11}=$20~K and $\Delta E_{11}=$1125~K for reaction~(\ref{f.o}), respectively.
The observed molecular abundances in our studies are indicated by the grey shaded regions, while the molecular abundances in TMC1 \citep{2013ChRv..113.8710A} are indicated by the cyan shaded regions. The other physical conditions are indicated on the top of this figure.
}\label{Fig:model-tk}}
\end{figure*}

In Fig.~\ref{Fig:model-tk}, we first compare the modeling results with $T_{\rm g}=$ 10~K and $T_{\rm g}=$ 20~K for a fixed H$_{2}$ number density of 10$^{4}$~cm$^{-3}$. These results suggest that the abundances change only slightly with gas temperature. This indicates that the HCNH$^{+}$ abundance does not significantly depend on the kinetic temperature in the range of 10--20~K. For HCNH$^{+}$, the abundances are slightly higher for $T_{\rm g}=$20~K than for for $T_{\rm g}=$10~K, which is different from Fig.~\ref{Fig:abun-tk}. This is likely because the difference caused by gas temperature variation is negligible when compared to the impact of H$_{2}$ number densities. Hence, we mainly focus on the dependence of chemical abundances on the H$_{2}$ number density. 

\begin{figure*}[!htbp]
\centering
\includegraphics[width = 1 \textwidth]{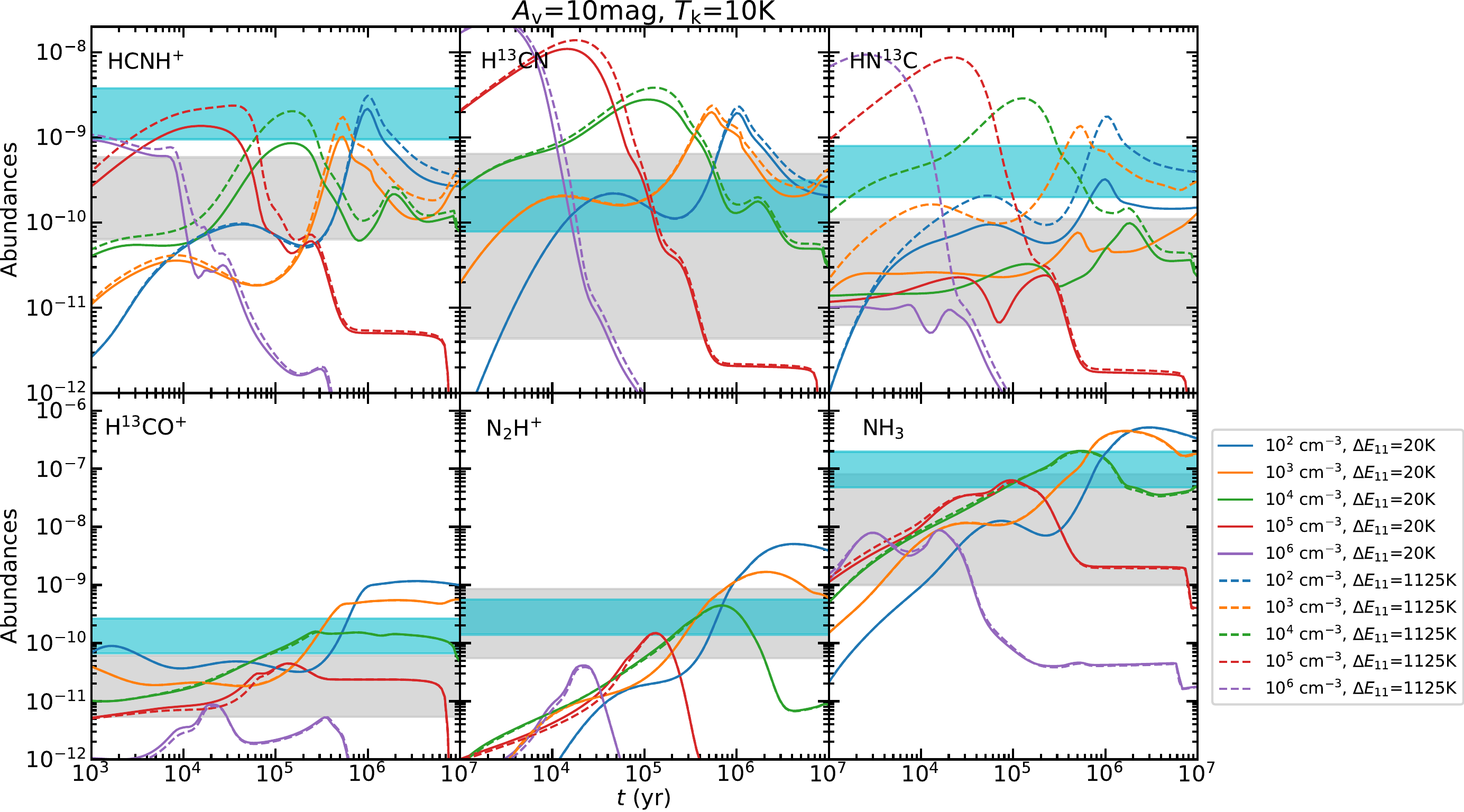}
\caption{{Molecular abundances relative to H$_{2}$ as a function of time, estimated from \textit{Chempl} \citep{2021RAA....21...77D}. The different colors correspond to the different H$_{2}$ number densities from 10$^{2}$~cm$^{-3}$ to 10$^{6}$~cm$^{-3}$. The solid and dashed lines represent the modeling results with the energy barrier $\Delta E_{11}=$20~K and $\Delta E_{11}=$1125~K for reaction~(\ref{f.o}), respectively.
The observed molecular abundances in our studies are indicated by the grey shaded regions, while the molecular abundances in TMC1 \citep{2013ChRv..113.8710A} are indicated by the cyan shaded regions. 
}\label{Fig:model}}
\end{figure*}

Figure~\ref{Fig:model} presents the time dependent chemical modeling results of the abundances of HCNH$^{+}$, H$^{13}$CN, HN$^{13}$C, H$^{13}$CO$^{+}$, N$_{2}$H$^{+}$ and NH$_{3}$, where a $^{12}$C/$^{13}$C isotopic ratio of 70 is adopted to obtain the abundances of the $^{13}$C-bearing molecules  \citep[e.g.,][]{1994ARA&A..32..191W,2016ApJ...824..136L,2023A&A...670A..98Y} and the ortho-to-para ratio of NH$_{3}$ is assumed to 4.0 \citep[see Fig. 3 in ][for instance]{2002PASJ...54..195T}. We find that the two different energy barriers for reaction~(\ref{f.o}) affect the results of HCNH$^{+}$, H$^{13}$CN, and HN$^{13}$C. The general trend is that assuming an energy barrier $\Delta E_{11}=$ 1125~K results in higher HCNH$^{+}$, H$^{13}$CN, and HN$^{13}$C abundances than assuming $\Delta E_{11}= $20~K. The HN$^{13}$C abundances are the most significantly affected. While our observed abundances can be roughly reproduced at an H$_{2}$ number density of $10^{3-4}$~cm$^{-3}$ and cloud ages of around $10^{5-6}$~yr in Fig.~\ref{Fig:model}, the high HN$^{13}$C abundances in TMC1 cannot be well reproduced by $\Delta E_{11}=$ 20~K suggested by \citet{2020A&A...635A...4H}. This indicates that the rate coefficients derived with $\Delta E_{11}=$ 20~K might not be suitable for the cold environments at $T_{\rm g}\sim$ 10~K. Therefore, we only use the results with $\Delta E_{2}=$ 1125~K suggested by \citet{2014ApJ...787...74G} for the following discussions. 

\begin{figure}[!htbp]
\centering
\includegraphics[width = 0.49 \textwidth]{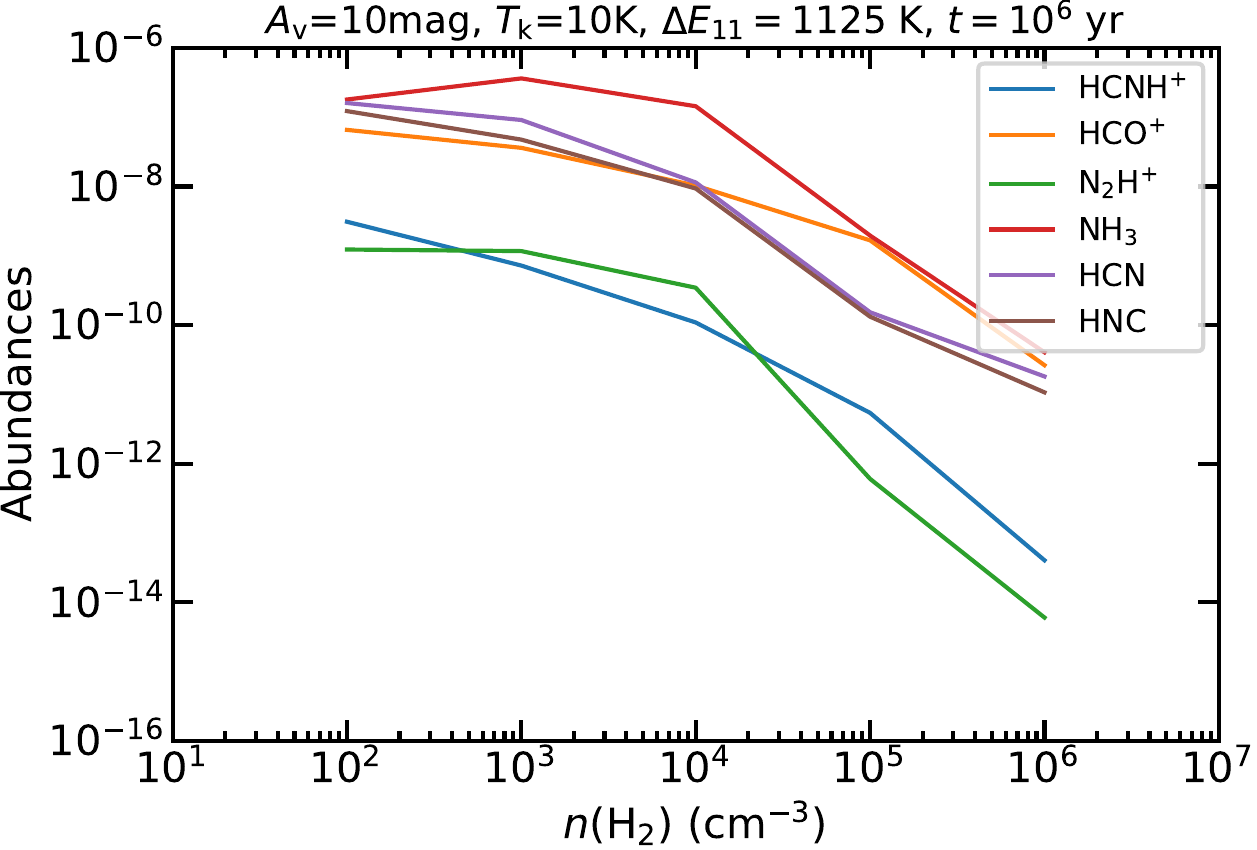}
\caption{{Molecular abundances of HCNH$^{+}$, HCN, HNC, HCO$^{+}$, N$_{2}$H$^{+}$, and NH$_{3}$ as a function of H$_{2}$ number density at a simulated timescale of $10^{6}$~yr. The adopted physical conditions are indicated on the top of this panel. 
}\label{Fig:density-abund}}
\end{figure}

\begin{figure}[!htbp]
\centering
\includegraphics[width = 0.49 \textwidth]{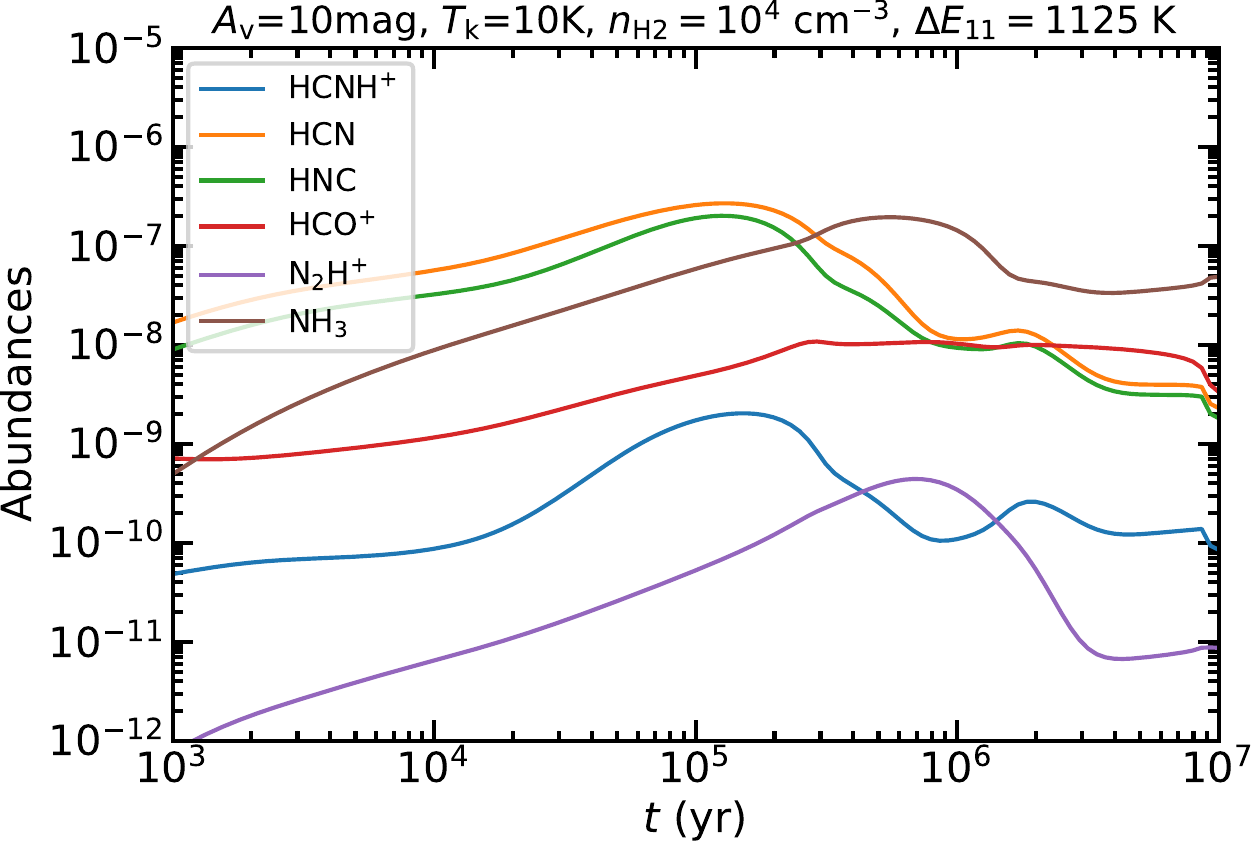}
\caption{{Temporal evolution of the abundances of HCNH$^{+}$, HCN, HNC, HCO$^{+}$, N$_{2}$H$^{+}$, and NH$_{3}$, estimated from \textit{Chempl} \citep{2021RAA....21...77D}. The adopted physical conditions are indicated on the top of this panel. 
}\label{Fig:freezeout}}
\end{figure}

In Figure~\ref{Fig:model}, we also find that the HCNH$^{+}$, H$^{13}$CN, and HN$^{13}$C abundances tend to reach the maximum earlier at higher H$_{2}$ densities and then decrease.
The timescales to reach the maximum are nearly identical for the three molecules, regardless of the H$_{2}$
number densities (see also Fig.~\ref{Fig:freezeout}), indicating that HCNH$^{+}$ abundances depend on the HCN and HNC abundances in the modeling results. Previous studies suggest that the reactions including Eqs.~(3)--(6) are the main formation path of HCNH$^{+}$
\citep[e.g.,][]{2014MNRAS.443..398L,2017MNRAS.470.3194Q,2021A&A...651A..94F}. 
HCN, HNC, H$_{2}$O, and HCO$^{+}$ are heavily depleted in cold and dense regions (see also Sect.~\ref{sec.col}).
Since HCN and HNC are thought to be the main precursors of HCN$^{+}$ and HNC$^{+}$ \citep[e.g.,][]{2014MNRAS.443..398L}, these cations are also depleted in these regions. Similarly, H$_{2}$O serves as the precursor for H$_{3}$O$^{+}$, which is also expected to be depleted under these conditions. If Eqs.~(3)--(6) are the main formation pathways of HCNH$^{+}$, this ion should also be depleted since its precursor molecules are heavily depleted. Therefore, the modeling results support that the decrease in its abundance is mainly caused by the freeze-out process, and the timescale of the abundance peak at a given H$_{2}$ number
density can be readily explained by the depletion timescale that inversely depends on the H$_{2}$ number density \citep[e.g.,][]{1999ApJ...523L.165C}. Figure~\ref{Fig:density-abund} presents the modeled abundances of HCNH$^{+}$, HCN, HNC, HCO$^{+}$, N$_{2}$H$^{+}$ and NH$_{3}$ as a function of H$_{2}$ number density at a simulated timescale of $10^{6}$~yr. This demonstrates the anti-correlation between the H$_{2}$ number density and molecular abundances, which readily explains the observed HCNH$^{+}$ abundance dependence of H$_{2}$ column density in Fig.~\ref{Fig:abun-h2}. These results suggest that HCNH$^{+}$ should be the most abundant in low-density regions where precursor species like HCN and HNC do not freeze out onto dust grains. Figure~\ref{Fig:freezeout} presents the molecular abundances reproduced by \textit{Chempl} as a function of simulation times for the same physical conditions (i.e., $T_{\rm g}$=10~K, $n_{\rm H_{2}}=10^{4}$~cm$^{-3}$). This figure supports the selective freeze-out scenario in which HCN and HNC deplete earlier than N$_{2}$H$^{+}$ and NH$_{3}$.

The modeling results are consistent with the observational finding that HCNH$^{+}$ is the most abundant in starless cores in their early evolutionary phase. Since the density is expected to increase during 
gravitational collapse, HCNH$^{+}$ abundances should decrease after the onset of infall. This also explains why the starless core TMC1 has a much higher HCNH$^{+}$ abundance than the prestellar core L1544 and our observed regions (see Fig.~\ref{Fig:abundance}). Investigation of HCNH$^{+}$ in different evolutionary stages of low-mass star formation (see Fig.~\ref{Fig:stage} and Appendix~\ref{app.evo}) and non-detection of HCNH$^{+}$ (3--2) around class 0 protostars \citep{2020A&A...635A.198B} are also in line with this scenario. On the other hand, mass accretion flows in SSN are indicated by previous observations \citep[e.g.,][]{2013MNRAS.436.1513F}. Hence, we suggest that the HCNH$^{+}$ abundance gradient along the SSN2 in Fig.~\ref{Fig:sers-abun}a can be a result of the longitudinal mass accretion. 

Although HCNH$^{+}$ depletion is indeed found and appears to correlate with HCN and HNC depletion in Sect.~\ref{sec.col} and Sect.~\ref{sec.dependence}, the HCNH$^{+}$ abundances in SSN1 show an anti-correlation with HCN and HNC, and a positive correlation with N$_{2}$H$^{+}$ and NH$_{3}$. Moreover, the HCNH$^{+}$ emission peak coincides with N$_{2}$H$^{+}$ emssion toward SSN1. These facts imply that HCNH$^{+}$ does not follow the depletion of HCN and HNC in SSN1. Such trends cannot be explained by our chemical modeling results (see Figs.~\ref{Fig:model} and \ref{Fig:density-abund}). This suggests additional HCNH$^{+}$ formation paths from molecules that do not freeze out. Because species like NH$_{3}$ and N$_{2}$ can still survive in the gas phase even when HCN and HNC are heavily depleted,  we suggest that the ion-neutral reactions from these species become more important in the formation of HCNH$^{+}$ toward freeze-out regions.



In order to further study the importance of the formation pathways for HCNH$^{+}$, we compare the rates of reactions (1)--(7). For a binary reaction, the reaction rate is defined as $v_{\rm rate}=kX_{\rm A}X_{\rm B}$, where $k$ is the reaction rate coefficient, and $X_{\rm A}$ and $X_{\rm B}$ are the abundances of the two reactants. However, the rate coefficients of reactions (1), (2) and (7) are not available in the RATE12 chemical network \citep{2013A&A...550A..36M}. The rate coefficients at 300~K were suggested to be 1.7$\times 10^{-9}$~cm$^{3}$~s$^{-1}$ and 7$\times 10^{-11}$~cm$^{3}$~s$^{-1}$ for reactions (1)--(2) \citep{1991A&A...247..487S}, respectively. In the RATE12 chemical network, the same paths to form its isomer H$_{2}$NC$^{+}$ exist with the rate coefficients of 1.5$\times 10^{-9}$~cm$^{3}$~s$^{-1}$ and 6.7$\times 10^{-11}$~cm$^{3}$~s$^{-1}$ at 10~K \citep{1977CPL....47..145S,1976ApJ...209..638F}. The rate coefficient of reaction~(1) agrees with the latest calculations within uncertainties \citep{2008ApJ...686.1486M}. Hence, these values are comparable to the values of \citet{1991A&A...247..487S}. This would be expected if the branching ratios of reactions to form HCNH$^{+}$ and H$_{2}$NC$^{+}$ are identical. 
Hence, we simply take the latter two values for reactions (1)--(2). Based on previous studies, the rate coefficient of reaction~(7) is 9$\times 10^{-10}$~cm$^{3}$~s$^{-1}$ at 10~K \citep{Knight1988,2014MNRAS.437..930L}. These values are thus used for the qualitative calculations. 

\begin{figure*}[!htbp]
\centering
\includegraphics[width = 0.95 \textwidth]{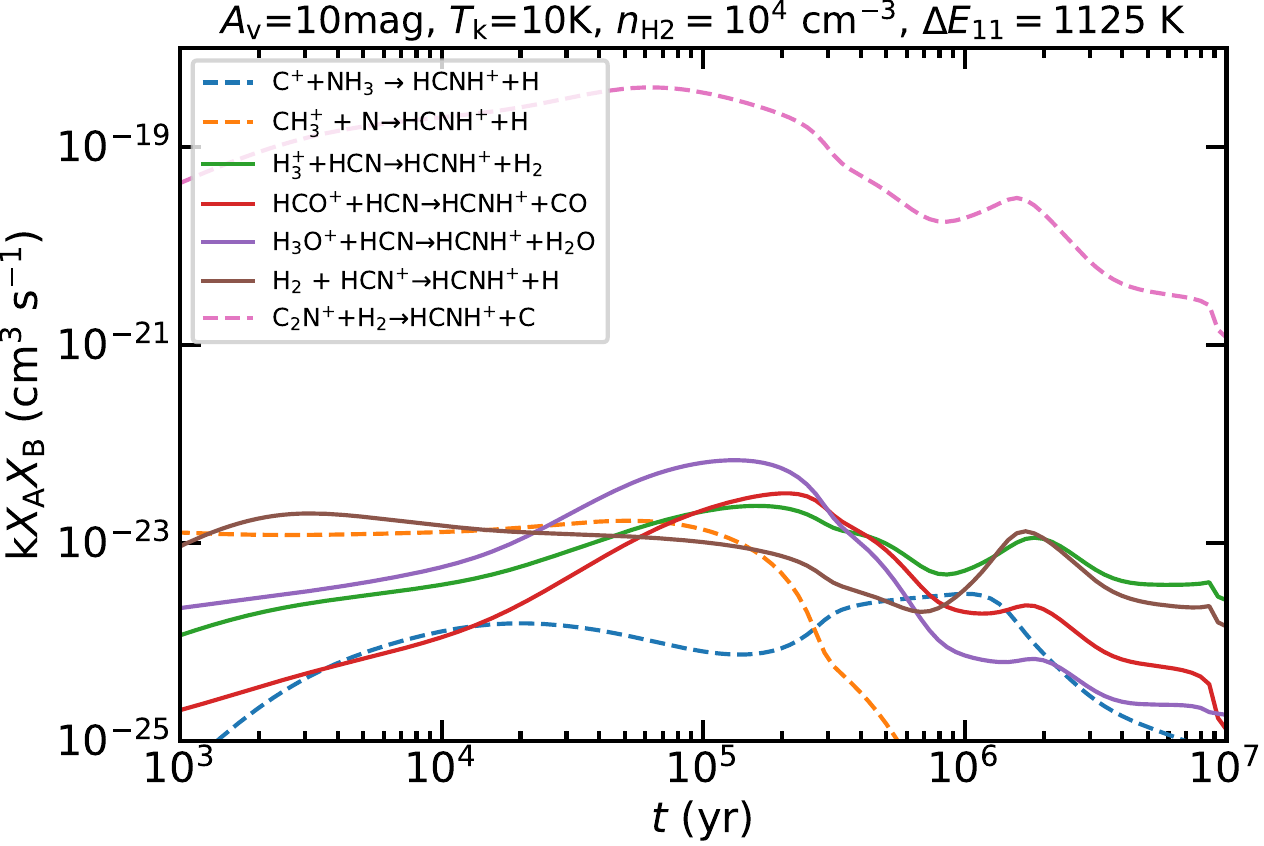}
\caption{{Reaction rates of the different formation paths of HCNH$^{+}$ as a function of time. The solid lines indicate reactions that are included in the RATE12 chemical network, whereas the dashed lines include potentially important reactions not incorporated in the network.
}\label{Fig:rate}}
\end{figure*}

Based on our chemical modeling results and the rate coefficients discussed above, we can estimate the rates of reactions~(1)--(7) to assess the relative importance of these formation paths as a function of time. Because reactions~(1), (2), and (7) are not included in our modeling calculations, we only took the molecular abundances from the modeling results to estimate the formation rates. Figure~\ref{Fig:rate} presents the comparison of the reaction rates for the chemical model using a number density of 10$^{4}$~cm$^{-3}$. We surprisingly find that reaction~(7) involving the hydrogenation of C$_{2}$N$^{+}$ emerges as the dominant formation pathway for HCNH$^{+}$, superseding all other competing reactions by several orders of magnitude. We investigate the modeling results with an updated chemical network augmented with reaction~(7) in Appendix~\ref{app.r7}, which suggests that reaction~(7) can make significant contributions to the formation of HCNH$^{+}$ at least on simulated timescales of $<10^{5}$~yr. This further casts doubts on the main formation path of HCNH$^{+}$ via reactions (1)--(6). Furthermore, reaction~(2) is as efficient as reactions (5)--(6) at timescales on the order of $\lesssim  10^{5}$~years. On the other hand, reaction~(1) is relatively less efficient, but becomes non-negligible at a timescale of $\sim 10^{6}$ years when compared with reactions (2)--(6). The significance of missing reactions, particularly reaction (7), in the formation of HCNH$^{+}$ is underscored by this comparison. More complete chemical networks and laboratory efforts to derive accurate rate coefficients will help to improve our understanding of the chemistry of HCNH$^{+}$ and eventually larger nitrogen-bearing molecules.




\subsection{Collisional excitation}
Owing to its small dipole moment \citep[0.29~D;][]{1986CPL...124..382B}, the low $J$ transitions of HCNH$^{+}$ have Einstein A coefficients of $<5\times 10^{-6}$~s$^{-1}$ (see Table~\ref{Tab:lin}), which are an order of magnitude lower than those of the other molecular transitions used in this work. By utilizing the latest rate coefficients governing collisions with H$_{2}$ \footnote{The adopted collisional rate coefficients are more accurate than those presented in \citet{2023JChPh.158g4304B}, because the $j=$2 excited energy level of H$_{2}$ has been taken into account in the latest results (Bop, C., priv. comm).} (Bop, C., priv. comm) and Eq.~(4) in \citet{2015PASP..127..299S}, we have calculated the optically thin critical densities of HCNH$^{+}$ (1--0), (2--1), (3--2) at 10~K to be 2.0$\times 10^{2}$~cm$^{-2}$, 1.7$\times 10^{3}$~cm$^{-2}$, and 5.2$\times 10^{3}$~cm$^{-2}$, respectively. Intriguingly, these critical densities are even lower than the corresponding low-$J$ CO critical densities in the optically thin regime \citep{2010ApJ...718.1062Y}, implying that the energy levels connecting these HCNH$^{+}$ transitions in LTE and can trace low-density regions. Therefore, we argue that HCNH$^{+}$ can be regarded as a good probe of pristine molecular gas prior to the onset of gravitational collapse. 

Recent studies have suggested that electrons may play a significant role in the excitation of molecular tracers like HCN and CH in molecular clouds \citep{2017ApJ...841...25G,2021A&A...650A.133J}. Adopting the same approximation that the collisional cross sections for electron excitation will be dominated by long-range forces and scale as the square of the permanent electric dipole moment \citep{2017ApJ...841...25G}, we can derive the HCNH$^{+}$--e$^{-}$ collisional rates by scaling the values of the HCN--e$^{-}$ and CH$^{+}$--e$^{-}$ collisional rates from \citet{2007MNRAS.382..840F} and \citet{2017MNRAS.469..612F}. Because the dipole moment (0.29~D) of HCNH$^{+}$ is about 0.10 and 0.17 times those of HCN \citep[2.985~D;][]{1984JChPh..80.3989E} and CH$^{+}$ \citep[1.683~D;][]{PhysRevA.75.012502}, the HCNH$^{+}$--e$^{-}$ collisional rates are found to be $\sim$2$\times 10^{-13}$~cm$^3$~s$^{-1}$ and $\sim$1$\times 10^{-12}$~cm$^3$~s$^{-1}$, respectively. Utilizing these values, we can estimate the critical electron fractional abundance that is required to make the electron collision rate equal to the H$_{2}$ collision rate, $x^{*}(e^{-})= n_{\rm c}(e^{-})/n_{\rm c}({\rm H}_{2})$ \citep{2017ApJ...841...25G}. $x^{*}(e^{-})$ is found to be $\sim(1-4)\times 10^{-3}$, which is considerably greater than the expected electron abundances of a few 10$^{-9}$ in molecular clouds \citep{2002ApJ...565..344C}. On the other hand, molecular clouds with H$_{2}$ number densities of $\gtrsim 5\times 10^{3}$~cm$^{-3}$ are sufficient to thermalize the excitation temperatures due to the low critical H$_{2}$ densities of the three low $J$ HCNH$^{+}$ transitions. Therefore, we conclude that electron collisions with HCNH$^{+}$ are unlikely to be significant in molecular clouds. We also note that the approximation of $x^{*}(e^{-})$ is very crude especially in the case of ion-electron interactions, because the dipole moment is no longer the only dominant term in the long-range forces. 

\section{Summary and outlook}\label{Sec:sum}
Using the IRAM-30 m and APEX-12 m telescopes, we have mapped HCNH$^{+}$, H$^{13}$CN, HN$^{13}$C, H$^{13}$CO$^{+}$, and N$_{2}$H$^{+}$ transitions toward the Serpens filament and Serpens South. Our primary findings are summarized below:

\begin{itemize}

\item[1.] Our observations of the two regions provide the first reliable distributions of HCNH$^{+}$, which suggests that HCNH$^{+}$ is abundant in cold and quiescent regions but shows a deficit toward star-forming regions (i.e., emb10, the Serpens South cluster). \\

\item[2.] An LTE analysis suggests that the observed HCNH$^{+}$ column densities range from $4.2\times 10^{12}$~cm$^{-2}$ to 2.7$\times 10^{13}$~cm$^{-2}$, which leads to its corresponding abundance relative to H$_{2}$ ranging from $3.1\times 10^{-11}$ in protostellar cores to $5.9\times 10^{-10}$ in prestellar cores. HCNH$^{+}$ is more abundant in molecular cores prior to gravitational collapse than in prestellar and star-forming cores. This result is in agreement with the scenario that HCNH$^{+}$ abundances decrease as the region evolves. \\

\item[3.] Based on our observations, we find that HCNH$^{+}$ abundances generally decrease with increasing H$_{2}$ column density. We suggest that HCNH$^{+}$ destruction in cold environments ($T_{\rm g}\sim$10~K) mainly depends on the increased H$_{2}$ number density which causes the freeze-out of its precursors HCN and HNC. \\

\item[4.] Current astrochemical models cannot explain the observed trend in SSN1 where the HCNH$^{+}$ abundance shows an anti-correlation with HCN and HNC, but shows a positive correlation with N$_{2}$H$^{+}$ and NH$_{3}$. This indicates that additional formation paths of HCNH$^{+}$ from molecules (e.g., N$_{2}$ and NH$_{3}$) that do not freeze out should play an important role in the formation of HCNH$^{+}$ in the freeze-out regions where HCN and HNC are heavily depleted. \\

\item[5.] Comparing the reaction rates of possible formation paths of HCNH$^{+}$, we find that important chemical reactions for the formation of HCNH$^{+}$ are likely missing in the RATE12 chemical network. More complete chemical networks and accurate rate coefficients are indispensable to understand the chemistry of HCNH$^{+}$. \\

\item[6.] The optically thin critical densities for HCNH$^{+}$ (1--0), (2--1), (3--2) at 10~K, resulting from collisions with H$_{2}$ molecules, are found to be 2.0$\times 10^{2}$~cm$^{-2}$, 1.7$\times 10^{3}$~cm$^{-2}$, and 5.2$\times 10^{3}$~cm$^{-2}$, respectively. These values suggest that LTE for these transitions should be readily achieved in molecular clouds. Since LTE is a good approximation for HCNH$^{+}$, the derived HCNH$^{+}$ column densities should be reliable (see Table~\ref{Tab:column}). On the other hand, electron excitation plays a negligible role for these transitions in molecular clouds. \\
 
\end{itemize}  

Our observations have shown that HCNH$^{+}$ appears to become deficient after the onset of gravitational collapse in low-mass star formation regions (see also Appendix~\ref{app.evo}). 
However, the observational picture toward other environments is still elusive. Although HCNH$^{+}$ has already been detected in a number of high-mass star formation regions \citep[e.g.,][]{2021A&A...651A..94F}, the influence of the environment on its spatial distribution is not well investigated in such environments. We also searched for information on HCNH$^{+}$ in previous line surveys of the circumstellar envelopes of evolved stars. HCNH$^{+}$ (2--1) has been observed in the 2 mm line survey of the asymptotic giant branch star IRC+10216 \citep{2000A&AS..142..181C,2008ApJS..177..275H} and HCNH$^{+}$ (3--2) has been covered in the 1 mm line survey of IRC+10216 and the red superginat VY CMa \citep{2010ApJS..190..348T,2013A&A...551A.113K}, but the HCNH$^{+}$ transitions were not detected by these sensitive surveys. This indicates that this molecular ion might not be abundant in circumstellar envelopes of evolved stars. For extra-galactic observations, this molecule appears to be tentatively detected in NGC~4945 \citep{2016cosp...41E2009V,2017IAUS..321..305V}.

\section*{ACKNOWLEDGMENTS}\label{sec.ack}
We acknowledge the IRAM-30 m and APEX-12 m staff for their assistance with our observations. Y.G. is grateful to Ashly Sebastine for her preliminary investigation of the APEX-12 m pointed observations toward Serpens South. Fujun Du is supported by the National Natural Science Foundation of China (NSFC) through grants 12041305 and 11873094. C.~H. has been funded by Chinese Academy of Sciences President's International Fellowship Initiative under Grant No. 2022VMA0018. A.M.J. acknowledges support by USRA through a grant for SOFIA Program 08-0038. X.D.T. acknowledges the support of the Natural Science Foundation of Xinjiang Uygur Autonomous Region under Grant No. 2022D01E06, the Tianshan Talent Program of Xinjiang Uygur Autonomous Region under Grant No. 2022TSYCLJ0005, and the Chinese Academy of Sciences "Light of West China" Program under Grant No. xbzg-zdsys-202212. C. Bop acknowledges financial support from the European Research Council (Consolidator Grant COLLEXISM, Grant Agreement No. 811363). M.R.R. is a Jansky Fellow of the National Radio Astronomy Observatory, USA. The research leading to these results has received funding from the European Union’s Horizon 2020 research and innovation program under grant agreement No 101004719 [Opticon RadioNet Pilot ORP]. This publication is based on data acquired with the Atacama Pathfinder Experiment (APEX). APEX is a collaboration between the Max-Planck-Institut f{\"u}r Radioastronomie, the European Southern Observatory, and the Onsala Space Observatory. This research has made use of NASA's Astrophysics Data System. This work also made use of Python libraries including Astropy\footnote{\url{https://www.astropy.org/}} \citep{2013A&A...558A..33A}, NumPy\footnote{\url{https://www.numpy.org/}} \citep{5725236}, SciPy\footnote{\url{https://www.scipy.org/}} \citep{jones2001scipy}, Matplotlib\footnote{\url{https://matplotlib.org/}} \citep{Hunter:2007}, APLpy \citep{2012ascl.soft08017R}, and seaborn\footnote{\url{https://seaborn.pydata.org/}} \citep{Waskom2021}. We would like to thank the anonymous referee for helpful comments.


\bibliographystyle{aa}
\bibliography{references}

\begin{appendix}
\section{Molecular column density maps}\label{app.ncol}
\begin{figure*}[!htbp]
\centering
\includegraphics[width = 0.9 \textwidth]{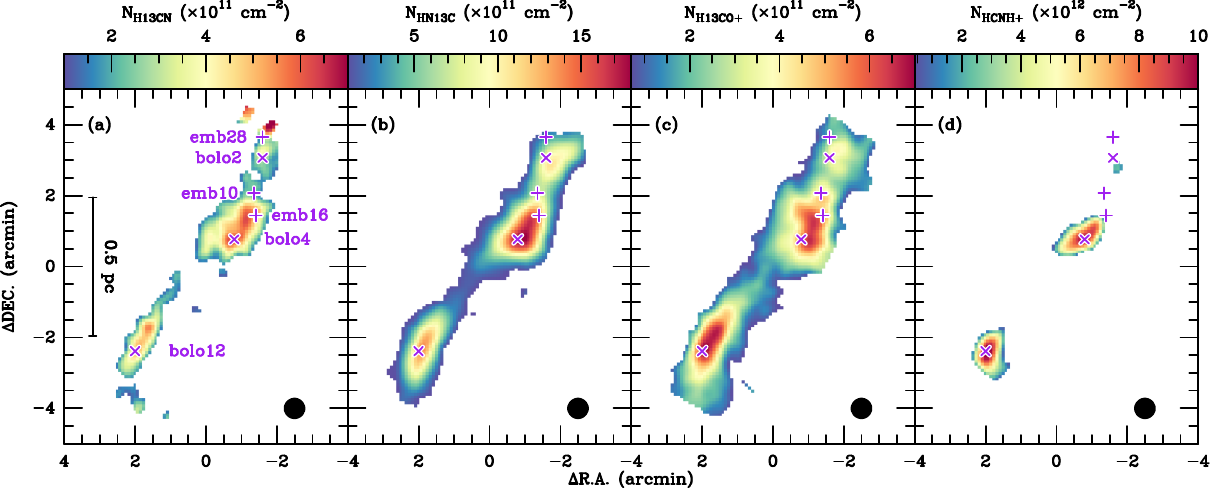}
\caption{{Molecular column density maps of the Serpens filament. The markers are the same as in Fig.~\ref{Fig:sf-hcnh+}.}\label{Fig:col-serf}}
\end{figure*}

\begin{figure*}[!htbp]
\centering
\includegraphics[width = 0.9 \textwidth]{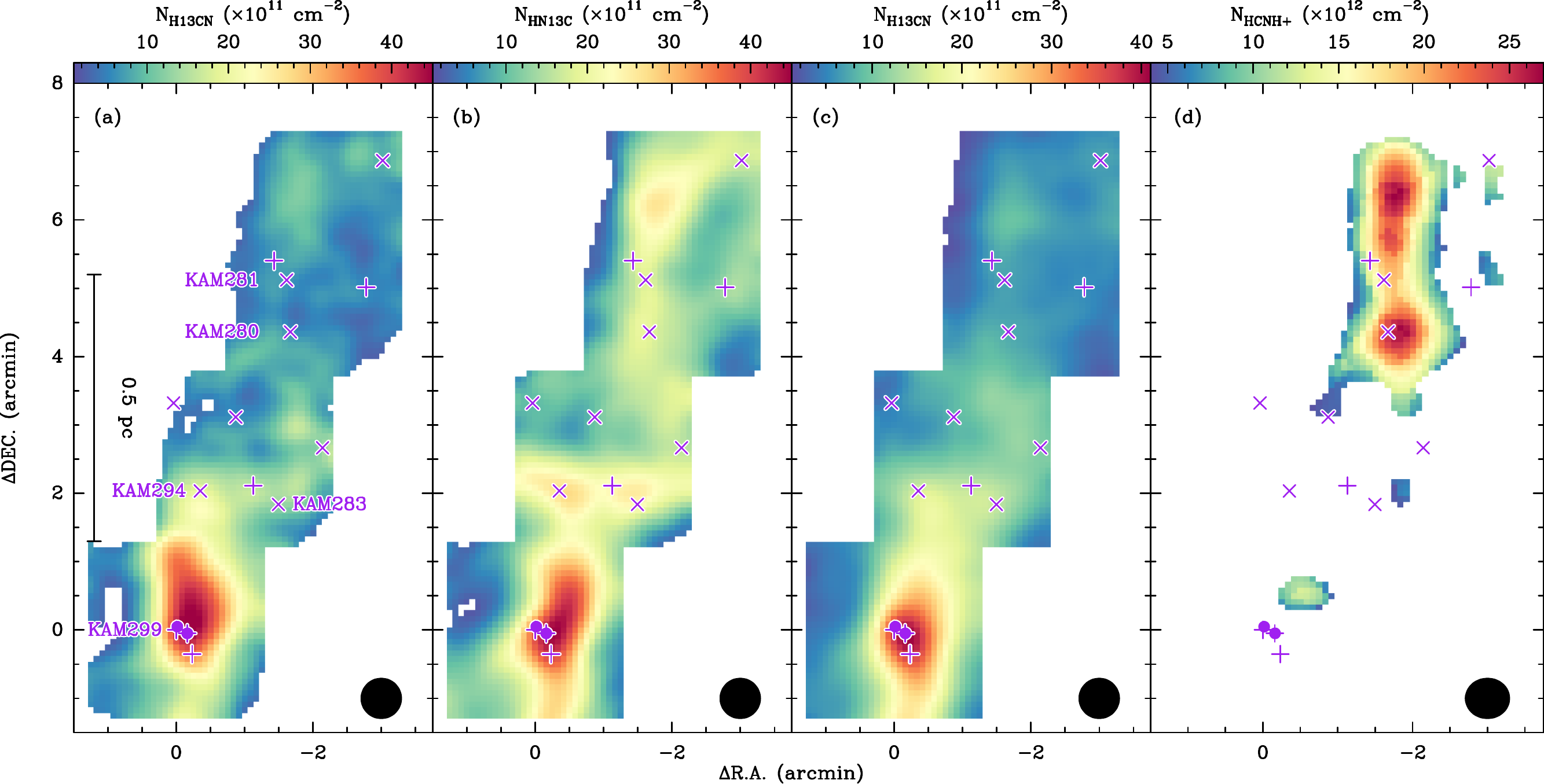}
\caption{{Molecular column density maps of Serpens South. The markers are the same as in Fig.~\ref{Fig:sers-hcnhp}.}\label{Fig:col-sers}}
\end{figure*}

We present the column density maps of H$^{13}$CN, HN$^{13}$C, H$^{13}$CO$^{+}$, and HCNH$^{+}$ in Figs.~\ref{Fig:col-serf}--\ref{Fig:col-sers}.

\section{HCNH$^{+}$ in different evolutionary stages of low-mass star formation}\label{app.evo}
\begin{figure*}[!htbp]
\centering
\includegraphics[width = 0.9 \textwidth]{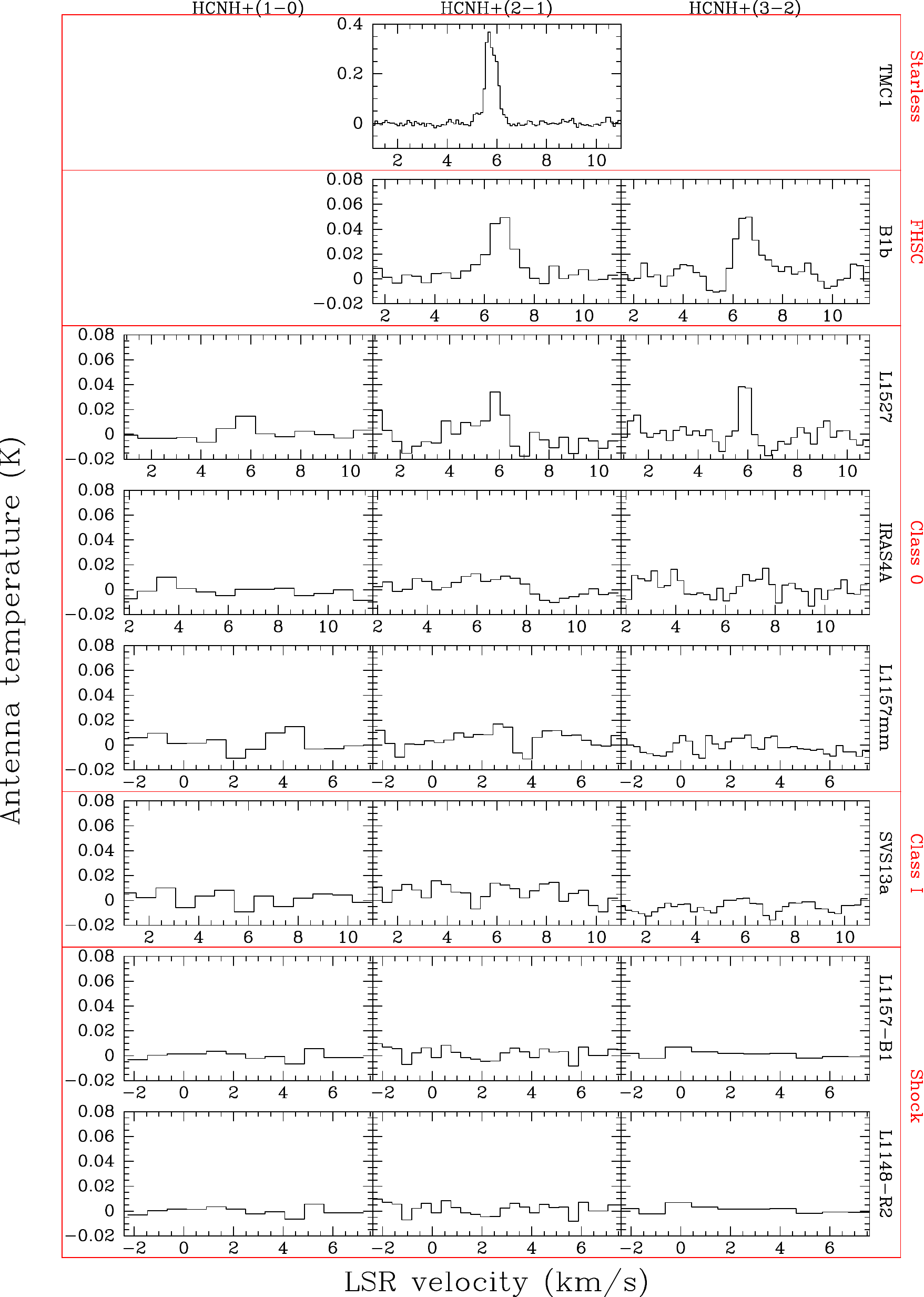}
\caption{{ASAI HCNH$^{+}$ spectra of TMC1, B1b, L1527, NGC1333-IRAS4A, L1157mm, NGC1333-SVS13a, L1157-B1, and L1148-R2 which represent different evolutionary stages, including starless cores, first hydrostatic cores (FHSCs), Class 0 protostars, Class I protostars, and shocked regions. HCNH$^{+}$ (1--0), (2--1), and (3--2) are shown in the first, second, and third columns. The source name is indicated on the right side.}\label{Fig:asai-hcnh+}}
\end{figure*}

Figure~\ref{Fig:asai-hcnh+} shows the HCNH$^{+}$ spectra of low-mass star-formation regions in different environments. The HCNH$^{+}$ spectra of TMC1, L1527, IRAS4A, L1157-mm, L1157-B1, SVS~13A, and L1448-R2 are based on IRAM-30 m observations, and the data are directly taken from the Large Program `Astrochemical Surveys At IRAM' \citep[ASAI\footnote{\url{https://www.iram.fr/ILPA/LP007/}},][]{2018MNRAS.477.4792L}. 
This figure reveals that HCNH$^{+}$ is only abundant in the early phases of low-mass star formation. During the Class 0 phase, L1527 is the only Class 0 object showing HCNH$^{+}$ emission. This trend is consistent with our finding that the HCNH$^{+}$ abundance decreases from the starless phase to the protostellar phase (see Fig.~\ref{Fig:stage}).


HCNH$^{+}$ is not detected in either the Class I object, SVS13A, or the shocked regions, L1157-B1 and L1148-R2. We also note that HCNH$^{+}$ was not detected in the line survey toward the protoplanetary disc, AB Aur \citep{2015A&A...578A..81P}. This indicates that HCNH$^{+}$ is not prominent in late stages of low-mass star formation. 


\section{Updated chemical models}\label{app.r7}
\begin{figure*}[!htbp]
\centering
\includegraphics[width = 0.95 \textwidth]{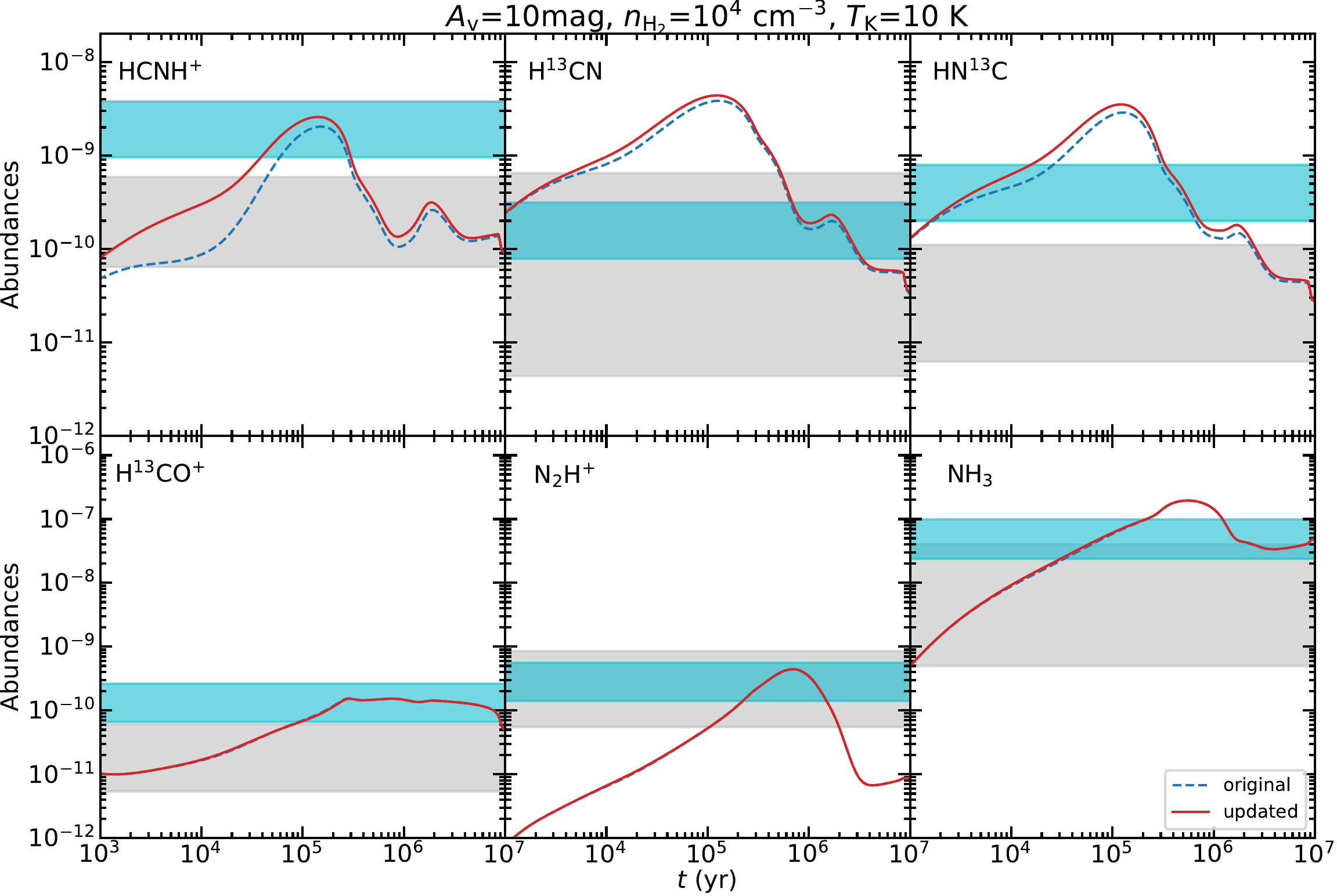}
\caption{{Molecular abundances relative to H$_{2}$ as a function of time, calculated from \textit{Chempl} \citep{2021RAA....21...77D}. The dashed blue lines represent the modeling results in Sect.~\ref{sec.chemistry}, while the  solid red curves represent the modeling results obtained when including the reaction~(7) to the underlying chemical network. The observed molecular abundances in our studies are indicated by the grey shaded regions, while the molecular abundances in TMC1 \citep{2013ChRv..113.8710A} are indicated by the cyan shaded regions. The other physical conditions are indicated on the top of this figure.
}\label{Fig:model-c2np}}
\end{figure*}

As demonstrated in Sect.~\ref{sec.chemistry}, reaction~(7) appears to play an important role in the formation of HCNH$^{+}$. Consequently, we have embarked on an exploration of chemical models, augmenting the existing chemical network with the inclusion of reaction~(7). The results of this endeavor are displayed in Fig.~\ref{Fig:model-c2np}, which showcases modeling results under fixed physical conditions with $A_{\rm V}$=10~mag, $T_{\rm g}$=10~K, and $n_{\rm H_{2}}=10^{4}$~cm$^{-3}$. Compared to the results with the chemical network without reaction (7), we find that HCNH$^{+}$, H$^{13}$CN, and HN$^{13}$C can reach higher abundances in the updated chemical models. Especially for a simulated timescale of $\lesssim 10^{5}$~yr, HCNH$^{+}$ abundances are much higher in the updated chemical models. The most substantial discrepancy emerges around 10$^{4}$~yr, exhibiting a factor of $\sim$5 difference. 

\begin{figure*}[!htbp]
\centering
\includegraphics[width = 0.95 \textwidth]{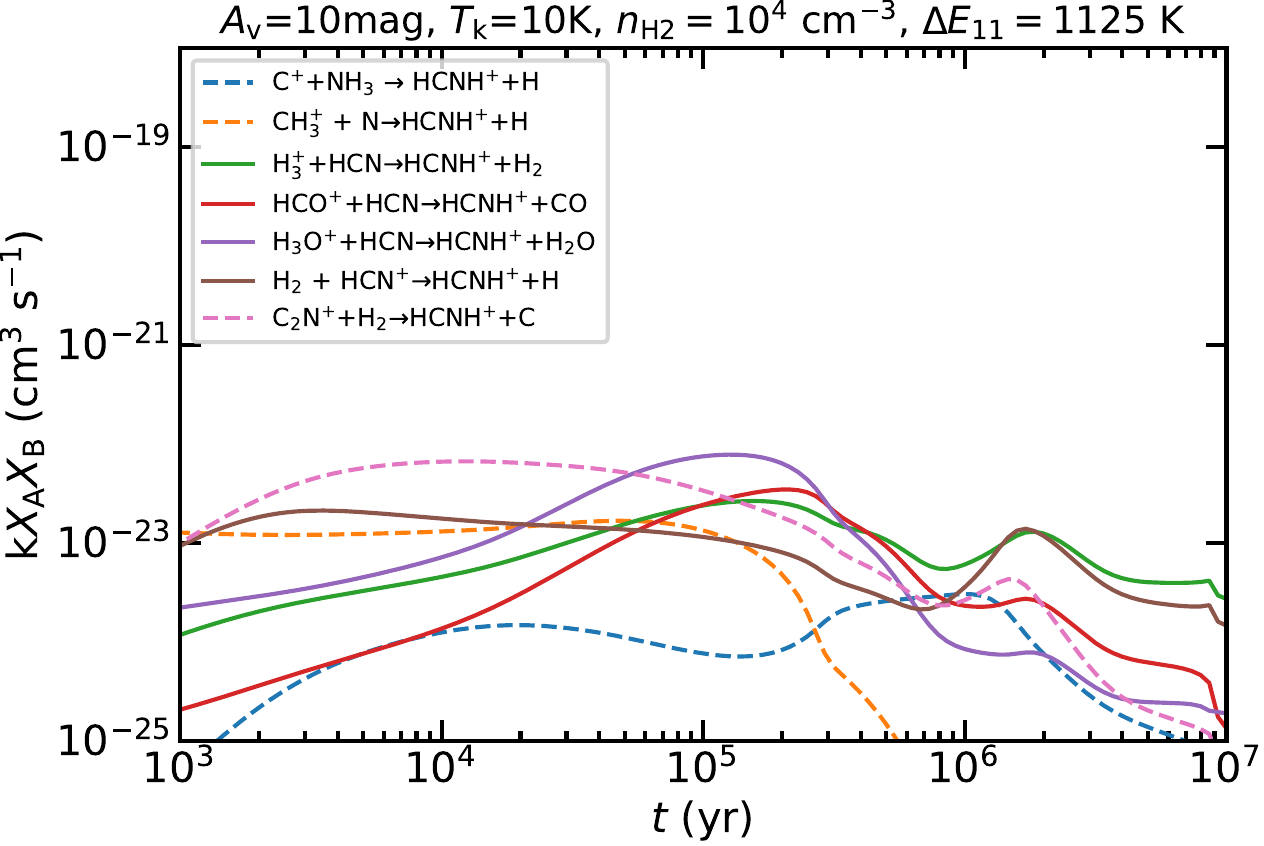}
\caption{{Same as Fig.~\ref{Fig:rate} but for including the reaction~(7) to the underlying chemical network. 
}\label{Fig:rate-c2np}}
\end{figure*}

We also revisit the formation rates with the updated chemical networks as in Fig.~\ref{Fig:rate}. The updated results are shown in Fig.~\ref{Fig:rate-c2np}. Compared to Fig.~\ref{Fig:rate}, the formation rate of reaction~(7)in the updated chemical model is significantly reduced by about four orders of magnitude. This is because reaction~(7) cause the rapid decrease of the C$_{2}$N$^{+}$ abundance in early time and the C$_{2}$N$^{+}$ abundance is about four orders of magnitude lower than in Fig.~\ref{Fig:rate}. It is still worth noting that the formation rate of reaction~(7) is still high before 10$^{5}$~yr when the freezeout process becomes more important. These facts demonstrate that reaction~(7) can make significant contributions to the formation of HCNH$^{+}$ at least at a simulated timescale of $<10^{5}$~yr. 

On the other hand, our findings regarding the overall abundance evolution remain remarkably consistent beyond 10$^{5}$~yr in Fig.~\ref{Fig:model-c2np}, because the freeze-out process becomes dominant in driving the abundance evolution. Because our targets are expected to have ages of $\gtrsim$~10$^{5}$~yr, the consistency between the models further strengthens the reliability of the results outlined in Sect.~\ref{sec.chemistry}. The abundances of H$^{13}$CO$^{+}$, N$_{2}$H$^{+}$, and NH$_{3}$ are identical across the simulated timescales from both sets of modeling results, implying that reaction~(7) plays a negligible role in determining the abundances of these particular species.

\end{appendix}

\end{document}